\documentclass[12pt,preprint]{aastex}

\newcommand{\cmcubed}{\hbox{${\rm cm^{\rm -3}}$}}
\newcommand{\kmps}{\hbox{${\rm km~s^{-1}}$}}
\newcommand{\Rsun}{\hbox{${R_{\sun}}$}}
\newcommand{\Msun}{\ensuremath{M_{\sun}}}
\newcommand{\Mdot}{\hbox{${M}\kern-0.6em^{^{\bullet}}$}}
\newcommand{\MLrate}{\hbox{$\Msun~{\rm yr}^{-1}$}}
\newcommand{\Ro}{\hbox{${R_{*}}$}}
\newcommand{\Rhox}{\hbox{$\int$\kern-0.05em$\rho$dx}}
\newcommand{\Texc}{\ensuremath{T_{\rm exc}}}
\newcommand{\Telect}{\ensuremath{T_{\rm e}}}
\newcommand{\nelect}{\ensuremath{n_{\rm e}}}
\newcommand{\nion}{\ensuremath{n_{\rm i}}}

\newcommand{\Hyd}[1]{\hbox{\ion{H}{#1}}}
\newcommand{\C}[1]{\hbox{\ion{C}{#1}}}

\newcommand{\Oxy}[1]{\hbox{\ion{O}{#1}}}
\newcommand{\Al}[1]{\hbox{\ion{Al}{#1}}}

\newcommand{\Mg}[1]{\hbox{\ion{Mg}{#1}}}
\newcommand{\Fe}[1]{\hbox{\ion{Fe}{#1}}}
\newcommand{\Si}[1]{\hbox{\ion{Si}{#1}}}
\newcommand{\pubdao}{Publ.\ Dominion Astrophys.\ Obs.\ Victoria}
\shorttitle{31~Cyg and Single Stars}
\shortauthors{Eaton}

\begin{document}

\title{A MODEL FOR THE CHROMOSPHERE/WIND OF 31~CYGNI \\
               AND ITS IMPLICATIONS FOR SINGLE STARS}

\author{Joel A. Eaton}
\affil{Center of Excellence in Information Systems, \\
       Tennessee State University,\\
       Nashville, TN}

\email{eaton@donne.tsuniv.edu}

\begin{abstract}
I develop a detailed empirical model for the chromosphere and wind of 31~Cyg 
based on a previously published analysis of {\it IUE} spectra from the 1993 
eclipse and on the thermodynamics of how the wind must be driven.  I then use 
this model to interpret observations of single supergiant stars and to assess 
the evidence that their winds are fundamentally different from those of 
supergiants in the binary systems.  This model naturally predicts a certain 
level of clumping of the gas to balance the pressure that drives the wind.  
It also predicts that anisotropic turbulence, such as would result from 
transverse displacements of Alfven waves directed along radial magnetic flux 
lines, would not give the roughly Gaussian profiles of emission lines seen 
in cool giant stars.  Furthermore, it implies that \C2] may not tell us much 
at all about general conditions in chromospheres.  Finally, I speculate that 
chaotic magnetic fields, in dynamical equilibrium with the gas of the wind, 
are the actual driving mechanism.
\end{abstract}


\section{INTRODUCTION}

It's no great stretch to assume that all cool stars (later than mid-F spectral 
type) have chromospheres and, likely, winds as well.  These are regions above the 
photosphere too hot to be heated radiatively; rather, they must be heated by magnetic 
processes or by shocks.  Most analyses such stars' chromospheres and winds are based 
on interpreting the strengths and shapes of emission lines in their spectra.  The 
Sun is the only one for which we can actually map out the chromosphere in any 
appreciable spatial detail.  All the others are unresolved, which limits us to 
interpreting global properties of the emitting gas in terms of models without any 
real spatial information (zero-dimensional).  There are a number of standard 
diagnostics used in this work.  For cool giant and supergiant stars, we have the 
\C2] intersystem multiplet at 2325 \AA\ (Stencel et al.\ 1981), the \Al2] multiplet 
at 2669 \AA\ (e.g., Judge 1986a, 1986b; Harper 1992), the \Mg2\ $h$ and $k$ lines 
(e.g., Stencel et al.\ 1980; Milkey et al.\ 1975), an abundance of \Fe2\ lines in 
the UV (ultraviolet) affected by scattering in winds (Judge \& Jordan 1991), 
H$\alpha$ (Johnson \& Klinglesmith 1965; Mallik 1993), and the \Hyd1\ Lyman lines 
(Landsman \& Simon 1993; Haisch et al.\ 1977).  Many of lines involved fall in the 
UV because of the temperatures and densities obtaining in these chromospheres, so 
our knowledge of cool giants' chromospheres necessarily has developed through 
observations from {\it IUE}, the {\it International Ultraviolet Explorer} satellite  
(e.g., Judge 1988) and more recently {\it HST}, the {\it Hubble Space Telescope} 
(e.g., Carpenter et al.\ cited below).  Other observations of these chromospheres 
involve measuring radio thermal emission from their winds (Drake \& Linsky 1986; 
Harper et al.\ 2001). 

An alternative, complementary approach to mapping the properties of chromospheres
solely by interpreting observations of the emission lines of single stars is to use
the $\zeta$~Aurigae binaries (Wright 1959, 1970; Wilson \& Abt 1954) to probe their
chromospheres and winds in greater spatial detail (legitimately at least 
one-dimensionally).  These systems consist of a cool supergiant paired with a B dwarf, 
the prototype system $\zeta$~Aur consisting of K4~Ib and B5~V components.  Absorptions 
in the spectrum of the B star by the gas in the wind and chromosphere of the K star give 
us a spatial probe of the K star's wind.  Again, the telling absorptions are in the 
UV, so most of our knowledge of these stars has come from {\it IUE} (e.g., 
Eaton 1996) with incremental improvements from {\it HST} (e.g., Baade et al.\ 1996).  
More recently Harper et al.\ (2005) have used radio observations of $\zeta$~Aur 
to verify the atmospheric structure they derive from UV lines.

The purpose of this paper is to develop a detailed physical model for the chromosphere/wind 
of 31~Cyg, to test it, and then use it to interpret observations of emission lines of single 
stars.  There is a commonly held belief, even prejudice, that the chromospheres/winds of 
single stars are different than the those of binary components in that winds of single stars 
accelerate much more rapidly than winds of the binaries and reach lower terminal velocities.  
This paper assesses whether the observations of single stars might be consistent with the 
slower acceleration of binary components.  I shall begin with a review of the picture 
developed for single stars through traditional analyses (\S\ 2), then discuss the model 
for cool supergiants developed from observations of the $\zeta$~Aur binary 31~Cyg (\S\ 3), 
testing it through comparisons with emission lines and radio continuum emission.  I then 
use this model to interpret the measurements for single stars (\S\ 4) and give some 
speculations about wind acceleration (\S\ 5).  My approach follows the radically different 
viewpoint that pressures supporting and accelerating the outer atmosphere are a 
{\it microscopic} phenomenon, operating on scales smaller than we resolve with our 
observations, and are not simply the manifestation of {\it global} waves.

\section{THE GENERAL PICTURE FOR SINGLE STARS}

Our observational understanding of single stars consists of several types of data.
(1) relative and absolute fluxes of a wide variety of UV and optical chromospheric 
emission lines, (2) profiles of optically-thin emission lines, reflecting the kinematics of 
gas forming them, (3) velocity structure of optically thick emission lines, reflecting bulk 
flows in the chromospheres/winds, (4) line strengths of density-sensitive multiplets, namely 
\C2] UV0.01 $\lambda$2325; and (5) strengths of fluorescent lines, reflecting mean intensities 
in some of the deeper parts of the chromosphere.  Many of these results come from {\it HST} 
spectra discussed in observational papers by K.G. Carpenter and his collaborators (Carpenter 
et al.\ 1991,1995,1997,1999; Robinson et al\ 1998); the more imaginative interpretations 
of them seem to have come mostly from P.G.\ Judge (e.g., Judge 1994).

The important results of these papers are (1) measurements of electron density in some part 
of the emitting region of the chromosphere, (2) turbulent velocities from optically-thin 
emission lines, (3) possible evidence for multiple components in the profiles of \C2], but 
not other optically-thin lines, as well as in the strengths of fluorescent molecular lines, 
(4) some indication of the temperatures in the chromospheres/winds from excitation of \Fe2\ 
and relative strengths of various emission lines, (5) detection of slight amounts of more 
highly ionized, possibly hotter, gas (e.g., \C4\ emission), and (6) evidence for atmospheric 
expansion and a crude measurement of the density-velocity profile from self-reversed \Fe2\ 
lines and \Mg2\ $h$ and $k$.

\paragraph{Chromospheric Velocity Structure:}  Many of the weaker intrinsic lines of common 
elements are weak enough in chromospheric spectra as to be optically thin.  They should be 
formed in essentially nebular conditions, with every excitation giving a potentially 
observable photon, so that their line profiles would reflect the kinematics and ionization 
structure of the emitting gas.  The superior resolution and signal to noise of {\it HST} 
have made it possible to measure their profiles reliably.  The turbulence measured from 
such an optically-thin line can be conveniently characterized by a velocity, $v_{0}$=FWHM/1.67, 
the parameter in the Gaussian part of the velocity distribution [exp$^{-(v/v_{0})^{2})}$], 
which Judge (1986a, 1986b) called $b$.  Carpenter et al.\ fit Gaussians to such lines in a 
variety of cool stars.  We may estimate the turbulence by averaging their results for 
roughly nine lines between 1900 and 2850 \AA, with the \C2] multiplet excluded, which give 
FWHM = 18.2 \kmps\ ($\alpha$~Tau), 29.9 ($\lambda$~Vel), 27.2 ($\alpha$~Ori), and 23.6 
($\gamma$~Cru).  I have corrected these values for the 13--15 \kmps\ resolution of the 
{\it GHRS} spectrograph which Carpenter et al.\ mention occasionally in passing, except in 
the cases they explicitly stated that they had corrected their published values.  These 
FWHM's correspond to turbulent velocities in the range $v_{0}$ = 10.9--17.9 \kmps\ for the 
emitting gas. This is well supersonic for the temperatures expected in these chromospheres 
($\lesssim$ 10,000~K). 

The most perplexing result of such kinematic analyses is the existence of broad wings and a 
$\sim$4 \kmps\ redshift of the \C2] lines in most of these stars.  These may well not be 
formed under similar conditions as most other weak lines, and, in fact, those other lines 
(such as fluorescent \Fe1\ and weak lines of \Fe2\ and \Al2]) do not show these phenomena.  
This discrepancy implies some sort of anisotropic turbulence seen only in \C2] or, perhaps, 
multiple components in the chromosphere, again not seen in the weak non-\C2] lines.  Single
Gaussians fit to the \C2] profiles in the various stars give values of FWHM = 24 \kmps\
($\alpha$~Tau), 28 ($\gamma$~Dra), 36 ($\lambda$~Vel), 35 ($\alpha$~Ori), and 30 ($\gamma$~Cru),
not corrected for the point spread function of the spectrograph.  For $\gamma$~Cru a two-component
fit gives FWHM=27/42 for core/wings; for $\alpha$~Ori, FWHM=19/48.  For $\lambda$~Vel, the star
most like 31~Cyg, we have a characteristic velocity of 21.6 \kmps, presumably uncorrected for
the resolution of the spectrograph, or 19.9 \kmps\ corrected; again, supersonic for the expected
chromospheric temperatures.  Of particular significance would be any differences of width in
these optically-thin lines; if the \C2] lines were formed primarily in a different region
from the others, they would not necessarily have the same broadening.  Indeed, the \C2]
lines are marginally broader than the other lines in $\alpha$~Tau ($\Delta$FWHM = 1.6 \kmps),
$\lambda$~Vel (3.4 \kmps), and $\gamma$~Cru (2.7 \kmps).  These differences must at least 
partially reflect the non-Gaussian profiles of \C2].

\paragraph{Electron Densities:} These come primarily from the density-sensitive line ratios 
of the \C2] (Stencel et al.\ 1981).  Electron densities derived from the \C2] line ratios in 
{\it HST} and {\it IUE} spectra of typical stars are near 10$^{9}$ \cmcubed.  It is important 
to remember that these values apply only to the parts of the chromosphere/wind where the photons 
are formed.  Because C is theoretically expected to be mostly neutral in the deeper reaches 
of the chromosphere, one might expect that the measured electron densities would apply only 
to the {\it warmer} parts of the chromosphere/wind.  The \C2] lines also give limits on the 
optical depths of chromospheric gas.

Judge (1994) analyzed the variations of very good observations of $\alpha$~Tau with Doppler 
shift to derive electron density as a function of velocity.  Again, his results were perplexing.  
They gave higher electron densities, by a factor of four, for positive velocities (away from us; 
ostensibly toward the star) than for negative velocities.  Furthermore, the emission in $\alpha$~Tau
was shifted to the red by 4 \kmps, which Judge interprets as a downflow in the denser regions 
of the chromosphere.  A redshift of 2--4 \kmps\ seems to be ubiquitous in the cool giants and 
supergiants (Judge \& Carpenter 1998).  On reflection, that the measured electron density is so 
similar in all these stars, in spite of variations in mass-loss rate of several orders of 
magnitude (see Harper 1996), is probably just as strange as the line shapes.

\paragraph{Line Strengths:}  Relative strengths of optically-thin lines give a measure
of the mass, temperature, and electron density of emitting gas averaged over the whole
chromosphere.  They may be interpreted with appropriate empirical or semi-empirical models 
as Judge (1986a, 1986b) did for several bright stars.  This sort of analysis provides 
global constraints on the integral
\begin{equation}
     f_{\rm line} \sim \int {\nion}{\nelect} exp(-\chi_{\rm exc}/k{\Telect})dV. \\
\end{equation}
Fluxes for characteristic lines come from {\it IUE} spectra and seem to be roughly consistent 
from star to star among the cool giants.  Such collisionally excited lines as Si~II] UV0.01, 
\Al2] UV1, and \Mg2\ $h$ and $k$ have consistent ratios for cool giants (Judge \& Jordan 1991).
\C2] UV0.01 is not necessarily proportional to the other collisionally excited lines
(Judge et al. 1992); however, it is roughly equal in strength to \Al2] for most of these 
stars.  For the values given by Judge \& Jordan, we can summarize these line strengths 
as in Table 1.  For the windy (non-coronal) giants, the flux in \C2] $\lambda$2325.4 seems to 
be roughly equal to flux in \Al2] $\lambda$2669.  We have also listed strengths for three 
$\zeta$~Aur binaries as best we can judge them.  The \Al2] fluxes for these binaries (Eaton 1992) 
are relatively well determined.  The \C2] fluxes, on the other hand, are very poorly measured 
and in a noisy part of the {\it IUE} spectrum.  There are no spectra for 31~Cyg exposed long 
enough to detect this line, and  it is hard to differentiate it from the noise in the spectra 
for $\zeta$~Aur and 32~Cyg in which Schr\"oder et al.\ (1988) claimed to detect it, although 
Harper et al.\ (2005) {\it may} have detected the strongest component with {\it HST}.  I have 
given what I consider the best measurements possible for these spectra, but in the following 
analysis, I shall assume f($\lambda$2325.4) $\approx$ f($\lambda$2669) in all the windy giants.  
The fluorescent \Fe1\ UV44 lines can be detected in all three classical $\zeta$~Aur systems 
(e.g., Bauer \& Stencel 1989), but these two lines, especially $\lambda$2844, are highly 
blended with other features.

Judge (1986a, 1986b; Judge \& Jordan 1991) has made simple empirical models for three cool giants, 
finding the emission is probably excited in a gas near 7000 K and an electron density of 10$^9$
(an assumed {\it global} value of \nelect\ from \C2] UV0.01).  Carpenter et al.\ (1999) estimated 
that the \Fe2\ lines observed in the wind of $\lambda$~Vel are scattered by gas at $\sim$ 6000 K 
from the relative strengths of various multiplets.  Other estimates of the electron temperatures 
of single stars come from adjusting semi-empirical models (e.g., Kelch et al.\ 1978) to give 
observed line profiles for charactistic emission lines and H$\alpha$ absorption.

\paragraph{Fluorescent Lines:} The classical fluorescent lines of K giants are \Fe1\ UV44 
excited by the \Mg2\ $k$ line.  These lines are optically thick and must be formed in the deeper, 
denser parts of the chromosphere for there to be any neutral iron to scatter them  (Harper 
1990).  Other fluorescent lines are scattered by molecular species (McMurray et al.\ 1999; 
McMurray \& Jordan 2000).  These molecular species must be present in some part of the chromosphere, 
but the eclipses of $\zeta$~Aur binaries generally do not detect them, probably because there is 
plenty of cold dense gas in the inner parts of the chromosphere to hide their absorptions in the 
competing atomic features.  McMurray et al.\ have argued that the strengths of fluorescent H$_2$ 
and CO lines, which they could not reproduce with their non-dimensional semi-empirical model of 
$\alpha$~Tau, require a multi-component atmosphere, possibly with shocks deep in the atmosphere 
to generate enough Ly$\alpha$ flux there to excite these molecules radiatively.  

\paragraph{More Highly Ionized Species:} Long exposures in the UV have detected lines of 
highly ionized species even in the windy giants.  These include intersystem lines of \Si3\ and 
\C3\ (e.g., Carpenter et al.\ 1999) and the resonance doublet of \C4\ (e.g., Robinson et al.\ 
1998).  In the Sun, such highly ionized species emit lines in the transition region at $\sim$ 
50,000~K.  In the windy giants, this interpretation is problematic because there is no evidence 
for the coronae that create the transition regions in the dwarfs.

\paragraph{Expansion-Velocity Structure of the Wind:}  This is very difficult to get from 
observations of single stars.  There is ample evidence of winds in the P-Cyg profiles of \Mg2\ 
and from the asymmetric, variable profiles of H$\alpha$ in cool supergiants (Zarro \& Rodgers 
1983; Mallik 1993).  Harper (1996) summarized analyses of wind profiles derived from single 
stars; Harper et al.\ (2005) have discussed this in somewhat more detail.  
Much of the evidence comes from P-Cyg profiles in the UV.  \Fe2\ lines of increasing 
intinsic strength show increasingly negative velocities of their self reversals, attributable 
to wind acceleration, in a number of cool supergiants, notably $\lambda$~Vel (Carpenter et al.\ 
1999), $\gamma$~Cru (Carpenter et al.\ (1995), $\alpha$~Ori (Carpenter \& Robinson 1997), and 
both $\alpha$~Tau and $\gamma$~Dra (Robinson et al.\ (1998).  Of these, $\lambda$~Vel had the 
most extensive coverage of the accelerating wind, with centers of the shell lines shifted 
to $\sim$ $-$32 \kmps\ with respect to the star, and Carpenter and Robinson have interpreted 
the data for these stars to infer mass-loss rates, terminal velocities, level of turbulence, 
and density structure.  Such analyses give winds accelerating much faster than those of the 
$\zeta$~Aur binaries or, for that matter, of $single$ stars as deduced from radio observations 
(Harper et al.\ 2001, 2005; Carpenter et al.\ 1999).  In fact, the analysis of thermal radio emission 
has become a fruitful technique for deducing the wind structure of single windy giants (e.g., 
Drake \& Linsky 1986; Harper et al.\ 2001).

\section{THE MODEL FOR 31~CYG}

Although optical spectra provided many insights into the nature of the extended atmospheres 
of the cool supergiant components of $\zeta$~Aur systems, the field was essentially dormant 
from the mid 1950's until {\it IUE} revitalized it with panchromatic UV spectra.  I have 
discussed the results of such studies in a review at the Cool Stars 9 meeting (Eaton 1996).  
I shall summarize them here as follows:  First, the new UV observations recorded absorptions 
from most of the important species expected to exist in chromospheres and winds of cool giant 
stars.  In contrast, optical spectra give very few of these species.  Furthermore these UV 
absorptions are often intrinsically strong lines that can be detected at great heights above 
the stellar surface, giving us the ability to probe winds to much greater height.  For the 
first time we could use the wings of Ly$\alpha$ to measure hydrogen column densities directly 
for many lines of sight through the wind/chromosphere.  The many lines of \Fe2, likewise, gave 
measurements of excitation temperature (T$_{\rm exc}$ $\sim$ 5,000--12,500 K) and kinematics 
of the wind throughout much of the wind and upper chromosphere.

Some of the results from {\it IUE} are conventional while others are surprising.  Strengths and 
shapes of lines from different ionization stages of metals, most importantly iron, confirm the 
expectation that the metals are mostly {\it singly-ionized} throughout the chromospheres and winds 
of these stars.  Likewise, the detection of the wings of Ly$\alpha$ at height in chromospheres and 
rough agreement of the mass column densities derived with those from \Fe2, means that {\it H 
is primarily neutral} throughout those parts of the wind we can sample.  This contradicts the 
predictions of semi-empirical models in which H becomes completely ionized over the 
first several scale heights of the chromosphere, giving roughly a constant electron density in the 
line-emitting regions in spite of a marked decrease of total density (Judge 1990, p.\ 290; \S\
3.2 below).  This obsevation that H remains neutral thus places limits on permitted kinetic 
temperatures and local electron densities in the gas.  Furthermore, wind models for 31~Cyg give us 
some direct insight into the turbulence in chromospheres.  Single-component models of the gas with 
no expansion effects, i.e., the sort of analyses used by Wilson \& Abt, find Doppler widths of the 
order of 20 \kmps, decidedly supersonic for gas with kinetic temperatures below 10,000~K.  
{\it IUE} observations show that at least some of this spread is caused by the differential 
expansion of the gas along the line of sight and need not be attributed to local turbulence.  In 
fact, the {\it IUE} observations for 31~Cyg require a turbulence $\lesssim$ 15 \kmps.

Even in the classical optical analyses, the ionization of metals was lower than expected for a 
uniformly distributed gas and thereby implied clumping for the gas to achive enough electron 
density to maintain an observable population of trace neutral species, such as \Fe1.   UV 
observations confirm this result.  Ionization throughout the wind is lower than expected from 
simplistic calculations of ionization equilibrium and implies clumping in the range  10--30$\times$ 
to achieve the inferred electron densities (n$_{e}$ $\sim$ 0.2--1.5 x 10$^9$ \cmcubed\ in the inner 
150 $\Rsun$ of the chromosphere).  This is a complication well beyond most semi-empirical models 
of chromospheres.

We can use an idealized description of the measured physical properties of the gas 
in the chromosphere and wind of 31~Cyg to test ideas about the structure of 
chromospheres and, ultimately, about wind mechanisms.  These measurements are based 
on the most extensive set of observations ever obtained for a $\zeta$~Aur binary 
(Eaton \& Bell 1994).  I have chosen to concentrate on 31~Cyg over the years because 
it has a much longer period than the other two classical $\zeta$~Aur binaries ($\zeta$~Aur 
and 32~Cyg).  This gives it a greater separation and less interaction between the wind 
and B star, a big advantage in many analyses.  The wind-scattered emission lines seen in 
total eclipse, for instance, are weaker than in the other two stars (Eaton 1992).  I am 
adopting the following properties for the 31~Cyg system: $D$ = 473 pc; $R_{\rm K}$ = 197 
$\Rsun$, $M_{\rm K}$ = 11.7 $\Msun$, $M_{\rm B}$ = 7.1 $\Msun$, $v_\infty$ = 90 \kmps, 
$\Mdot$\ = 3.0$\times$10$^{-8}$ \MLrate\ (Eaton 1993c; Eaton \& Bell 1994).  We should 
note that, in observing the more complicated $\zeta$~Aur, Baade et al.\ (1996) probably 
had a better case than 31~Cyg for applying their wind-scattering analysis.  Furthermore, 
inasmuch as the models for 31~Cyg and $\zeta$~Aur give essentially the same results in 
terms of velocity structure/extension of the wind and wind temperatures, we can be 
confident that the results for either of them should be readily applicable to other stars.

Table 2 gives the details of the empirical chromospheric/wind structure I am using
to test wind models.  Quantities listed are (1) radius (distance from center of the 
star in $\Rsun$), (2) expansion velocity, (3) excitation temperature, (4) a temperature 
to drive the wind thermally, (5) a clumping factor, CF, giving the inverse of the fraction 
of space actually filled with matter, (6) log($n_{\rm H}$), the total hydrogen density, 
(7) log(\nelect), the local electron density from an assumed constant ionization of H 
(3\%) and amount of clumping, (8) log(\nelect) for the variable ionization developed in 
\S\ 3.2 and with the assumed clumping, but limited to 10\% in the outer chromosphere, 
and (9) $v_{\rm equ}$, a velocity derived from equipartition between mass motion and internal 
thermal energy of the gas (see \S\ 5.2).  

Although the electron densities measured from photoionization balance imply significant 
clumping of the gas, they are rather crude and may be systematically wrong.  An independent 
way to estimate this degree of clumping is to look at the difference between the 
pressures\footnote{Strictly speaking, the wind is driven through the pressure {\it gradient}, 
which appears in the hydrodynamical momentum equation (\ensuremath{F = ma}), but a given 
pressure distribution, the higher level abstraction I am specifying, implies a pressure 
gradient, even in the simplistic case of an isothermal wind.} required to drive the observed 
wind and the gas pressures that would be available in the wind if it were not clumped.  This 
approach works because the gas pressures ($\sim$ $\rho$T$_{\rm gas}$/$\mu$) would be in 
equilibrium with whatever pressure drives the wind.  We do this by determining a temperature 
structure to drive the observed wind {\it thermally}, as though these stars had a coronal, 
i.e., generalized Parker-type, wind (see Lamers \& Cassinelli 1999, Chapters 4\&5), then 
comparing those temperatures to the observed temperatures, point by point, assuming the 
excitation temperatures we have measured approximate the electron temperature.  This last 
assumption seems reasonable because the excitation temperatures are similar to temperatures 
derived for semi-empirical chromospheric models of single stars, at least in the inner parts 
of the wind, but we will test it in later sections.  To reiterate, if the gas is clumped, 
thereby bifurcating into dense clumps and an unspecified interclump medium, there must be a 
pressure in the interclump medium maintaining the clumping.  We will assume that this is the 
pressure driving the wind and that it somehow breaks the wind up into microscopic clumps and 
maintains them.  This is fundamentally different from the approach of traditional wave models 
in which the gas is uniform and the waves act on it continuously.

The temperature structure to produce the pressure gradient required to drive the wind of 31~Cyg 
thermally, which we have derived from a thermal-wind model, is given roughly by the equations
\begin{equation}
     T_{\rm therm} = 20,000 K + 75,000 K (r - \Ro)/\Ro\  \hskip30pt  for \ \ (r-\Ro) < 150 \Rsun, \eqnum{2a} \\
\end{equation}
and \\
\begin{equation}
     T_{\rm therm} = 95,000 K - 75,000 K (r - 3\Ro)/(15\Ro) \hskip30pt for \ \ (r-\Ro) > 2\Ro,  \eqnum{2b} \\
\end{equation}
\setcounter{equation}{2}
where $T_{\rm therm}$ is thermal (electron) temperature for a uniformly distributed gas, 
r is the radius, and $\Ro$=197 $\Rsun$\ is the photospheric radius of the star.
I don't claim that this structure is a rigorous determination of the thermal-pressure profile,
but that it is adequate for giving a good idea of the pressures required throughout the
wind.  Figure 1 shows the velocity structure calculated for this adopted temperature
structure of Eq.\ 2; Figure 2, the amount of clumping implied.

A way to test this sort of model is to calculate the emission expected from it.  I have 
done this in two ways, first, with a spherically symmetrical model that calculates the line 
strengths, line profiles, and chromospheric mass column densities specifically for 31~Cyg.  
The second way uses a traditional plane-parallel model for $\zeta$~Aur I developed in 
1992--1995 with PANDORA (Vernazza et al.\ 1973, Avrett \& Loeser 1992) to explore the 
ionization structure and effect of clumping in these stars.

\subsection{Tests of a Spherically Symmetrical Model for 31~Cyg}

The spherical model for 31~Cyg has the distribution of gas given in Table 2.  The velocity 
structure and distribution of mass density (given as n$_{\rm H}$=$\rho$/1.4m$_{\rm H}$) come 
from fitting spectra of shell lines in atmospheric-eclipse spectra; they should be fairly reliable.  
Excitation temperature comes from fitting the excitation of \Fe2\ in these atmospheric-eclipse 
spectra; they should be reliably measured but subject to uncertainty about their meaning.  
Electron densities depend on level of ionization of H and degree of clumping.  We can 
make educated speculations about these properties as follows.  Since H seems 
to be neutral observationally, we might expect the ionization to be $\lesssim$ 10\%.
We will assume it constant with height, following the thoughts of Judge (1990, \S\ IIIb), 
and of the order of 1--5\%, and adopt a value of 3\% for the sake of a first-order model.  
In that case, the gas must be clumped even to approach the canonical \C2] electron density 
almost anywhere in the chromosphere.

\subsubsection{Line Emission}

For optically-thin lines, we can calculate the emission with the standard physics given
by Osterbrock (1974), for example, with assumptions about ionization and chemical
abundances.  In this approximation, emissivity [ergs~cm$^{-3}$s$^{-1}$] is just 
\begin{equation}
     \epsilon_{\nu} =  8.63\times10^{-6} \Omega{_{1,2}} n{_i} n{_e} exp(-\chi/kT) T^{-0.5} \omega{_1}^{-1} \chi  \\
\end{equation}
where n${_i}$ is the density of emitting ions, $\Omega$ is the collision strength, $\chi$ is
the excitation energy of the upper level, and $\omega{_1}$ is the statistical weight of the
lower level.  In this approximation, emission is proportional to collision strength, and we
have incorporated collision strengths for \C2] $\lambda$2325 ($\Omega$ = 0.830 and 1.66, for
transitions out of the two ground-state levels, with half the emission going into the dominant
2325.4\AA\ line) from Blum \& Pradhan (1992) and for \Al2] $\lambda$2669 ($\Omega$ = 3.3)
from Tayal et al.\ (1984).  Abundances of C and Al were $A_{\rm C}/A_{\rm H}$ = 2.0 
$\times$ 10$^{-4}$ and  $A_{\rm Al}/A_{\rm H}$ = 3.0 $\times$ 10$^{-6}$ (Eaton \& Bell 1994, 
\S\ 2.1).  To determine the total emission in an optically thin line as a function of radius 
on the sky, we sum the emissivity along a ray through the atmosphere, with integration limits 
determined by whether or not it intersects the star.  To get a profile for the line, we define 
a velocity scale and map the emissivity profile of the gas at each point in the atmosphere onto 
it, with allowance for the atmospheric expansion, $v_{\rm exp}$, an arbitrary (systematic) 
velocity along the line from the center of the star, $v_{\rm syst}$, and a spectrum of Gaussian 
turbulence, $v_0$.

The first thing we note from these models is the effect of the extended spherical atmosphere 
on the emission strength and line profiles.  Optically thin lines formed in such a structure 
will be highly {\it limb brightened} and will combine contributions from the face of the star 
and from a halo beyond its limb in similar proportions.  Figure 3a illustrates this effect for 
our 31~Cyg model for an emitter assumed to exist in the same ionization stage throughout the 
wind (\nelect\ as in Table 2, Column 7).  The emission peaks somewhat more to the limb than mass 
because electron density drops rapidly with height in these atmospheres, even with the assumed 
clumping.  An emitter that is depleted in the inner chromosphere, as \C2] may be, would peak even 
further beyond the limb.  This is shown in the figure as a dashed line calculated for emission 
only at temperatures in the model above 5200~K.  Even so, most of the emission would come from 
within the first $\lesssim$ 0.15~$\Ro$ of the chromosphere/wind.  Strong resonance lines, on 
the other hand, should be {\it limb darkened}.  Scattering in an extended atmosphere would give 
a highly forward-peaked source function (e.g., Cassinelli \& Hummer 1971), the reason for the 
core-halo profile of the discs of WR stars, and, for resonance lines in chromospheres, the 
large optical depths mean the escape probability for photons migrating through the damping 
wings would be much smaller for the radial direction than for other lines of sight.

Simple calculations like those in Figure 3a cannot represent the conditions in actual chromospheres 
because they give {\it fluxes} much higher than observed.  Table 1 gives fluxes calculated for three 
cases, (1) uniform 3\% ionization of H, (2) variable ionization of H, limited to 10\%, and 
(3) variable ionization of H, limited to 3\%.  The model with uniform H ionization 
arbitrarily set at 3\% gives an \Al2] flux too high by a factor of ten and a \C2] flux at least as
bad.  The model with variable ionization (Table 2, Column 8), on the other hand, gives \Al2] flux 
high by $\sim$50\% and \C2] high by slightly more, but the high electron density in the outer 
winds of these models, combined with single ionization of metals, gives much larger line broadening 
from differential expansion of the atmosphere.  These models also have problems with electron density.
If we average \nelect\ over emission of \C2], the values are either too high, as for the calculation for 
the unrealistic uniform 3\% ionization of H, or somewhat low, as for the realistic ionization of 
H limited to 3\% or even 10\%.  Figure 3b shows the intensity profiles for the model with variable 
ionization.  The effect of the 10\% ionization in the wind is obvious, with much of the emission in 
the wind coming from the hot ionized outer parts.  These parts, however, have electron densities much 
too low (7.7 in the log) to give the \C2] line ratios, at least the ones seen in single stars, and 
they produce \C2] profiles much broader than observed.  They are clearly inappropriate for single 
stars unless C becomes {\it doubly ionized} in the outer wind or unless the temperature in the outer 
wind is well below the measured excitation temperatures.

\subsubsection{Radio Continuum Emission}

A further test of these spherical models comes from the radio continuum observations of Drake 
\& Linsky (1986) and Drake et al.\ (1987), who detected a number of bright cool supergiants 
at the 0.1 mJy level (or 10$^{-27}$ ergs~cm$^{-2}$s$^{-1}$Hz$^{-1}$).  Their measurement for 
31~Cyg (0.36 $\pm$ 0.07 mJy) gives a check on the consistency of the electron densities and 
temperatures we are assuming.  The basic idea (see Harper et al.\ 2005) is to integrate the 
radio source function, $S_\nu$ = $B_\nu$, along rays through the wind and then sum the resulting 
intensities over the stellar disc to get a kind of luminosity [ergs~s$^{-1}$~ster$^{-1}$Hz$^{-1}$].  
This quantity would then be converted to flux at the Earth by multiplying it by the solid angle 
of 1~cm$^2$ at the star's distance, $D$. The dominant free-free opacity from electron-proton 
interactions, corrected for stimulated emission, is just 
\begin{equation}
     \kappa_{\nu} =  0.212 n{_p} n{_e} T_{e}^{-1.35} \nu^{-2.10} \hskip20pt  \\
\end{equation}
(Harper et al. 2005, Eqn.\ 3), while the generally smaller contribution from neutral-hydrogen--electron 
interactions (H$^-$) is roughly 
\begin{equation}
     \kappa_{\nu} =  6.2\times10^{-35} n{_H} n{_e} (\lambda/1 cm){^2} T_{e} \hskip20pt  \\
\end{equation} 
(Allen 1973, \S\ 42 evaluated at 8000 K).  For the values of \Telect,CF,\nelect\ in Table 2 
(Column 8 for \nelect), we calculate a flux at the Earth of 0.04 mJy at 6.2 cm, only about 10\% 
of the amount measured by Drake et al.  Figure 4 shows the calculated intensity profile, a slightly 
limb-brightened disc with a radius of $\sim$ 3.8 $\Ro$.  This failure to reproduce the 
flux seems to be a common problem in reconciling the radio emission of cool giants with their 
UV spectra.  Carpenter et al.\ (1999), for instance, were unable to reproduce the 3.5-cm 
flux of $\lambda$~Vel with the rapid wind acceleration they found from UV lines.  Harper et al.\ 
(2005), likewise, were unable to reproduce the radio spectrum of $\zeta$~Aur with their model and 
found it hard even to get the flux level.

To get emission as great as is observed for 31~Cyg, the outer wind must be essentially wholly 
ionized, as the large mass-loss rate of Drake et al.\ implies.  If we let the outer wind become 
mostly ionized, we can at least approach the flux level observed.   For instance, with H in the 
envelope 75\% ionized above 8000~K, we get a flux of 0.22 mJy; for 100\% ionization, 0.28 mJy. 
Is this level of ionization reasonable?  It seems consistent with observations of $\zeta$~Aur 
stars, especially 31~Cyg, but the evidence for single stars is ambiguous.  Such high ionization 
should reveal itself through absorptions of {\it doubly} ionized species in spectra of $\zeta$~Aur 
binaries and possibly through emission lines from these species in both the binaries and single 
supergiants.  The broadening of these lines would be a  measurement of the terminal velocity and 
turbulence of the outer wind.  Let us look at the observations of these species in some actual 
stars.  There is a weak emission feature at the position of \Si3] $\lambda$1892 in the eclipse 
spectra of 31~Cyg ({\it IUE} images SWP47335 and SWP47336).  In addition, strong emission lines 
of higher ionization are seen in eclipse in all three classical $\zeta$~Aur binaries.  Multiplets 
\Al3\ UV1 and \Fe3\ UV34 in 31~Cyg have P-Cyg profiles (Bauer \& Stencel 1989, Fig.\ 3) and are so 
strong they must be formed by scattering of light from the B star in the wind.  The line widths 
at the bases of their profiles are roughly 170 \kmps, about what one would expect for a wind at 
terminal velocity.  Eaton \& Bell found that these absorptions are formed primarily at velocities 
{\it toward the B star} and likely result from ionization of the outer wind by the B star, not by 
intrinsic radiation of the supergiant.  Thus a large fraction of the outer wind is ionized 
{\it in the 31~Cyg system}.  Since H requires less energy to ionize than the 19~eV to ionize 
C$^+$ out of its metastable level, H is likely ionized, as well, in these regions of 31~Cyg.
 
Single stars may have lower ionization in their outer winds than the binaries with similar mass-loss 
rates, although there is evidence in P-Cyg profiles of wind lines that some of the metals are doubly 
ionized in many of them (see \S 4.2 below).  Carpenter et al.\ (1999) detected lines of \C3] and 
\Si3] in $\lambda$~Vel with about the right broadening and strength to be formed in an ionized wind.  
However, they interpreted these lines as an indication of gas at $\sim$ 50,000~K, as though high 
ionization necessarily means high temperature.  The problem with wind formation is that the computed 
line profiles are wrong.  The calculated profiles are essentially square, having a central dip 
reflecting the absence of gas at low expansion velocities, while the observed profiles, although 
somewhat noisy, seem to have a central peak as though formed at least partly in turbulent gas at 
low expansion velocity.  Therefore the {\it emission} lines of single stars unfortunately tell us 
nothing about the ionization of their outer winds.

\subsection{Calculations with PANDORA}

PANDORA is useful for exploring non-traditional models of chromospheres because it lets one
specify an arbitrary distribution of turbulence to increase the scale height in hydrostatic
equilibrium and to specify an arbitrary distribution of added electron density that can be used
to simulate clumping.  I have made exploratory calculations for three cases: (1) a chromosphere
for $\epsilon$~Gem (G8~Ib) based on the model of Basri et al.\ (1981), (2) a generalized model 
of a $\zeta$~Aur binary ($\Ro$=150 $\Rsun$) with temperatures and scale heights based on my analyses
of $\zeta$~Aur (Eaton 1993a) and 32~Cyg (Eaton 1993b), and (3) models for $\alpha$~Tau to explore
excitation of H$\alpha$ in cool giants (Eaton 1995).  In the paper about H$\alpha$, I explored
the conditions necessary for exciting H$\alpha$ and \C2] in semi-empirical hydrostatic models,
finding that a certain amount of clumping of the warmer gas was required to strengthen H$\alpha$
and to have electron densities high enough to give reasonable \C2] line ratios.

These semi-empirical models can give us some insight into how the gas might be ionized at 
various depths in actual chromospheres.  All the models, regardless of assumptions about 
scale height, clumping, and temperature, predict species such as Al will be singly ionized 
throughout the whole emitting atmosphere.  For H and C, the calculated ionization increases 
with height.  Figure 5 shows this effect for H.  In it, I have 
parameterized height with the electron temperature, since, by assumption, temperature 
increases monotonically with height above a temperature minimum in all such semi-empirical 
models.  The calculations show that log(n$_{\rm e}$/n$_{\rm H}$) increases from a minimum 
of $-$4.0 at $\sim$ 2600~K, determined by ionization of the metals included in the calculation, 
to 0.0 (complete ionization of H) at $\sim$ 9400~K.  For C, complete ionization occurs 
near 5000~K in these models.  The actual models have scale heights somewhat smaller than 
that of 31~Cyg, so the level of ionization could be higher in 31~Cyg.

\subsection{A Further Question about the 31~Cyg Model}

One possible error in Eaton \& Bell's analysis of 31~Cyg is the determination of the temperature 
from excitation of \Fe2.  We have assumed that the excitation temperatures measured are {\it bona 
fide} electron temperatures of the gas.  This might be the case, in that most of these lines arise 
from excited metastable levels which would probably be in thermal equilibrium with the electrons, 
at least in the denser parts of the wind (e.g., Judge et al.\ 1992).  Harper et al.\ (2005) argued 
this point explicitly, while others deriving temperatures for $\zeta$~Aur binaries (such as Eaton 
\& Bell) have implicitly assumed it.  However, the excitation temperatures in the outer parts of 
the winds may well be greater than the thermal temperatures, since these regions are bathed in the 
radiation of a B star, and since the electron densities expected in these zones are lower than the 
critical density ($\sim$ 10$^6$ per Judge et al.\ 1992) for radiative processes to become important.
In the model for 31~Cyg, these temperatures have very little effect on the emission at any wavelength,
since the densities are very low.

Contrariwise, all of these stars have rather high excitation temperatures in their outer
chromospheres, regardless of the effective temperature of the B companion.  We see this in Figure 6 
which gives \Texc\ as a function of tangential mass column density through the chromosphere.  We 
see that the temperature rises to about 8500~K where the Balmer lines become optically thin.  This 
agrees roughly with theoretical calculations for supergiants.

\section{IMPLICATIONS FOR {\it SINGLE} STARS}

Here I shall attempt to fit the rich
lore of space-dimensionless analyses of line emission and absorption of single stars
into the context of our knowledge of the legitimately one- to two-dimensional knowledge
of $\zeta$~Aur binaries.  We might expect the wind structure of the binaries to be 
essentially the same as that of single stars because the strengths of intrinsic 
emissions formed in the wind and chromosphere of $\zeta$~Aur binaries are similar to 
those of single stars (Schr\"oder et al.\ 1988; Eaton 1992; Harper et al.\ 2005, Fig.\ 6), 
and the H$\alpha$ profiles, and their variation, seem to be the same as well (Eaton 1995; 
Eaton \& Henry 1996).

\subsection{The Reality of Semi-Empirical Models}

Most detailed analyses of cool stars' chromospheres are based on semi-empirical models in which 
one posits a temperature rise through the outer atmosphere, calculates a density structure from
hydrostatic equilibrium and radiative transfer, and then calculates the emergent spectrum from the 
physical conditions derived (e.g., Kelch et al.\ 1978; Basri et al.\ 1981; McMurray 1999).  These 
models are inspired by the solar chromosphere, in which there is a temperature rise from a minimum, 
determined by the location where non-radiative heating begins to dominate heating by photospheric 
radiation, to a point where the chromosphere merges with a transition region of rapidly increasing 
temperature, heated by downward conduction from a corona.  The increase in temperature with height 
follows naturally from observations of the Sun and from the physical consideration that temperature 
should rise with decreasing density to keep emissivity from falling precipitately with height per 
Equation (1).  Of course, there might be other ways to organize a chromosphere.  For example, the 
material at different temperatures might be more intimately mixed throughout the chromosphere, as 
McMurry et al.\ (1999) contemplate to some degree.  Harper et al.\ (2001) likewise have speculated 
about a truely radical reinterpretation of $\alpha$~Ori in which the chromospheric line emission 
comes entirely from hot inclusions in a generally cool neutral wind, although Harper et al.\ (2005) 
did not attempt to apply this radical approach to $\zeta$~Aur.  Indeed, temperature measurements 
for $\zeta$~Aur binaries, at least to first order, confirm some sort of general temperature rise 
with height.  Figure 6 shows the relation between excitation temperature and tangential mass column 
density through the atmospheres for four stars.  The temperature rise is obvious.  While this 
evidence does not preclude mixing some hot gas with the bulk of the warm gas throughout the 
chromosphere, it shows that the bulk of the gas behaves sort of like the gas in these classical 
semi-empirical models.

Another fundamental property of semi-empirical models is the density structure.  In the calculated 
models, it usually comes from hydrostatic equilibrium between gravity and thermal motions of the 
emitting gas ($T$ $<$ 10,000~K).    This is problematic in that there are obviously other sources of 
momentum flux in a typical chromosphere, such as the turbulence we see in profiles of emission lines, 
which will extend the atmosphere, changing its mass and electron density structure.  Furthermore, the 
mere existence of turbulence implies some sort of clumping of the gas, which would necessarily change 
the local electron densities and ionization structure.  Moreover, such effects would change the 
transfer of radiation and escape of photons from the chromosphere.

A third property of many semi-empirical models is a precipitate temperature rise to coronal 
temperatures defining the top of the chromosphere.  This is not a necessary feature, especially 
in the windy giants.  Nevertheless, McMurry (1999) used such a rise for $\alpha$~Tau to fit the 
Ly$\alpha$ profile and calculate emission from highly ionized species.  However, the observations 
of $\zeta$~Aur binaries do not require such a rise and may not even allow it; the profiles of \C3] 
and \Si3] seem to require formation at least partly in the inner chromosphere/wind, and the 
arguments of McMurry et al.\ (1999, 2000) suggest there are other places to excite \C4.

\subsection{Terminal Velocities of the Winds}

One of the basic tenets of our understanding of the windy giants is the idea that their terminal 
velocities are much lower than the surface escape velocity (e.g., Hartmann \& McGregor 1980; 
Judge 1992; Harper 1996).  These terminal velocities are probably not as well known as we think 
they are, and they may well be {\it much} higher than generally thought, especially in the normal 
K giants like $\alpha$~Tau, as Judge (1992) argued in a provocative paper about wind energetics. 
This is hardly a new idea (see, e.g., Judge 1992; Ahmad \& Stencel 1988; Kuin \& Ahmad 1989), but 
it is certainly worth discussing further in the context of supposed differences between single stars 
and binary components.  For the classical $\zeta$~Aur binaries, we base the terminal velocity on 
measurements of shell lines seen at all phases (e.g., Eaton 1993c).  We should be suspicious of low 
terminal velocities since the metals in the outer parts of the winds might well be doubly ionized, 
as they are in calculations of Harper et al.\ (2005, Fig.\ 5) for $\zeta$~Aur and \S\ 3.2 above 
for single stars.  Furthermore, the recombination time for H, given the electron densities in 
the outer parts of our model ($\sim$ 10$^{3-5}$ \cmcubed), is of the order of 1--100 yr.  
Semi-empirical models for single stars also become fully ionized, at least in H, in these 
regions, if that actually means anything.  So we really don't know what these winds are doing in a 
vast volume of space before they recombine and form the shell lines.  Occasionally more of the 
velocity structure may reveal itself through abnormally dense winds, as in AL~Vel in 1992 (Eaton 
1994, Fig.\ 8) and in $\lambda$~Vel in 1990 (Mullan et al.\ 1998).  Furthermore, we should realize 
that the winds must be sweeping up interstellar gas in these shells (cf. Wareing, Zijlstra, \& 
O'Brien 2007).  This means that the shell velocities are, if anything, likely to be {\it lower} 
than the true terminal velocities.

Carpenter et al.\ (1999) admit that we probably do not see much of the velocity structure in the 
traditional shell lines in many stars, while they contend that they have seen it all in $\lambda$~Vel.  
However, it is not clear from the line profiles of \Mg2, \Oxy1\ UV2, and \Fe2\ UV1 that one sees it 
even in that case.  The velocity structure for $\lambda$~Vel in 1994 seems to be ionization-bound, 
with the maximum wind velocity sampled limited by ionization of Fe$^+$ to Fe$^{+2}$.  In an 
ionization-bound wind, the relatively high density behind the ionization front would allow somewhat 
weaker lines to become thick at their normal rate, while the strongest lines would quickly saturate 
at the velocity of the ionization front, little more than the velocity of those somewhat weaker 
lines, to give the sort of levelling off seen in the highest measured velocities.  The edge velocities 
for all these strong lines in 1994 are around 50 \kmps, much greater than the supposed terminal 
velocity of 33 \kmps.  Furthermore, in 1990 the star showed absorptions to at least 80 \kmps\ 
(Mullan et al.\ 1998), which places its terminal velocity close to what we think 31~Cyg has, if the 
increase really did come from lower ionization in the outer wind.  Also, in comparing $\alpha$~Tau and 
$\gamma$~Dra, Robinson et al.\ (1998) found terminal velocities of 30 and 67 \kmps, respectively, 
for stars that otherwise seem to have similar atmospheric structure.  All this evidence suggests 
terminal velocities several times as large as they often seem, with the {\it apparent} terminal 
velocity dependent on just how much material is being ejected into the flow at a given time.  

There is another problem in interpreting these shell lines (cf.\ edge velocities of $\lambda$~Vel) 
that would lead to errors in the terminal velocity.  Carpenter et al.\ have interpreted the 
widths of Gaussians fit to the shell lines as a measure of {\it turbulence}.  This could conceivably 
be the case for \Fe2, but it is clearly inappropriate for at least some lines, for instance, for \Mg2\ 
in $\gamma$~Cru (Carpenter et al.\ 1995; Fig.\ 10), where \Mg2\ seems to have a significantly lower 
expansion velocity than even moderate \Fe2\ lines.  Such corrections for wind turbulence seem wrong, 
from both observational and theoretical considerations.  Observationally, in models for scattering 
in shells (e.g., Baade et al.\ 1996), the turbulence drops with height.  Theoretically, one would 
expect the turbulent energy to go into accelerating or heating the wind and be essentially damped 
out by the time the wind reaches its terminal velocity.

\subsection{Turbulence of the Chromospheric Gas}

To investigate the nature of the turbulence, we may calculate the profiles of broadening for 
various velocity distributions for the model of 31~Cyg.  The best fits seem to be for {\it isotropic} 
turbulence.  In Figure 7, the dashed line shows the profile for isotropic turbulence giving roughly 
the line broadening of single supergiants.  A Gaussian fits it quite well, as expected for the 
assumptions in the calculation.  The solid curve, on the other hand, shows the effect of imposing 
a $v_{\rm syst}$=10~\kmps\ downward velocity on this same turbulence.  The profile is still Gaussian 
at the level of the plot, but it is shifted to the red (3.8 \kmps) and broadened slightly more 
($\Delta$FWHM = 1.0 \kmps) by the variable projection of the systematic velocity into the line of sight.  

Calculations for {\it anisotropic} turbulence all give non-Gaussian profiles to some extent.  
Some of these are shown in Figure 8.  Radial turbulence is especially bad in this regard 
(Figs.\ 8a,c) because the strongest emitting regions are near the stellar limb, where the 
turbulent motions would be mostly perpendicular to the line of sight.  Tangential turbulence is 
considerably better but still decidedly non-Gaussian (Figs.\ 8b,d), with enhanced wings not seen 
in profiles of actual stars (see, e.g., Carpenter et al.\ 1991, Fig.\ 3b).  Figures 8c and 8d, 
however, show that {\it elliptical} anisotropic turbulence, in which one component is much larger 
than the other, is considerably closer to the observed Gaussian shape, and the departures from 
the observed Gaussian profiles would be somewhat reduced by convolving the calculated profiles 
with a Gaussian point-spread function for a spectrograph, which I have not done.  However, 
these profiles would still have overly broad wings not seen in actual stars.  I chose the nature 
of elliptical turbulence to reflect an isotropic turbulence of 5 \kmps\ from perhaps thermal 
motion and 20 \kmps\ of either radial or tangential macro/micro-turbulent motions.

Carpenter \& Robinson (1997) and Robinson et al.\ (1998) have cleverly fit the broadened wings of 
\C2] lines with an anisotropic turbulence somehow averaged over a uniform disc, but that model is 
clearly inappropriate in light of the large limb effects predicted by $\zeta$~Aur binaries (Figure 3).
In fact, a calculation with our model, but limited to gas within the limb of the star, gives the awful 
non-Gaussian profiles seen in Figure 8e.  

Given the distribution of mass in the model and the resulting limb brightening, the Gaussian 
profiles of intrinsic lines imply a fairly {\it isotropic} turbulence.  They certainly do not 
allow mostly {\it radial} turbulence, and they make it unlikely that the turbulence is 
perponderantly tangential.

\subsection{Electron Density and Interpretation of the \C2] Multiplet}

We have hardly any actual knowledge of the electron densities in the winds/chromospheres of any 
of these stars.  What we do know is how dense the regions emitting the bulk of \C2] are.  
Everything else must somehow come from a {\it model} calculation.

One of the triumphs of applying physics to cool stars is the use of \C2] to derive precise 
electron densities for a large number of objects.  The kinematic analysis of Judge (1994), 
however, undermines the importance of this result.  The redshift he detects in most stars 
is difficult to reconcile with other aspects of UV analyses, since it requires a 
$\sim$ 10~\kmps\ downdraft of all the emitting gas or something more extreme if only part 
of the gas is participating.  It would be convenient if this redshift were the result of a 
systematic error in the energy levels, and this may ultimately prove to be the cause of it.  
The energies of the upper levels of \C2] are quoted by Moore (1970) to the nearest 0.1 cm$^{-1}$, 
about 1 \kmps, while the wavelengths are somehow known to one more significant figure.  Likewise, 
the enhanced electron densities derived for the red side of these profiles seem to manifest 
themselves in only one of the line ratios, as though it might be affected by an unrecognized blend, 
but the dominant 2325.4\AA\ line would have to be the one affected.  

However, if we accept the reality of the shift and enhancement, \C2] must be telling us something 
profound about wind acceleration.  C is two orders of magnitude more plentiful than Al, yet lines 
of the two are roughly the same strength in cool giants.  Very little of the C, therefore, may be 
emitting \C2], and it must be produced under rather special conditions to give the observed 
multiplet ratios of single stars.  The broad wings of \C2] can be fit by emission in the 
expanding gas of the upper chromosphere/lower wind with the broadening produced by differential 
projection of expansion into the line of sight.  However, for that approach to work, the 
inner regions of the chromosphere must be neutral in C, even more so than implied by ionization 
balance in semi-empirical models such as those of \S\ 3.2 or measured in 31~Cyg.  Figure 8f shows 
such a calculation in which the gas below 5200~K emits no \C2].  This calculation gives a reasonable 
flux, but it does not fit any of the other properties of \C2].  The average electron density in the 
emitting gas (5$\times$10$^{7}$) is nowhere near the canonical 10$^{9}$, {\it even with the clumping}.  
The width of the profile is somewhat more than expected for a real star (FWHM = 35 vs.\ 30 \kmps\ 
for a weak line like \Al2]), the peak is shifted blueward by $-$0.8 \kmps, and the wings are probably 
too weak relative to the core.  Clearly a more radical departure from the expected ionization/velocity 
structure is required.

Looking at this more radically, we can get an idea of the conditions required {\it in the 31~Cyg 
model} by extending our assumptions about ionization.  Given the velocity structure, observationally 
all the C would be inert below 7000--8000~K to weight the emission at significant expansion velocity 
enough to give the extended wings (giving the same sort of effect as the anisotropic turbulence 
dismissed in \S\ 4.3), with the bulk of emission coming from locally dense, probably rather cool 
regions at depth that may have peculiar velocities.  Again, this means that C would be {\it much 
less ionized} in these atmospheres than predicted by semi-empirical models, with the \C2] emission 
coming from only a moderately small fraction of the gas.  However, the similarity of Doppler widths 
of \C2] and other optically-thin lines such as \Si2] and \Al2] suggests that both are formed in gas 
with similar turbulence, perhaps in similar parts of the atmosphere.  A possible source of the redshift 
of the line core is preferential excitation in the downward-facing parts of downward-moving elements 
in the turbulence distribution.  This fraction of the gas would have the greatest relative speed 
with respect to the outward momentum source and might well be expected to be subjected to the most 
violent accelerations.  Figures 9 show two experiments in simulating this effect.  In Figure 9a the 
emission comes from two sources, all the gas above 8500~K emitting with the velocity structure of 
Table 2 and 30~\% of the gas at temperatures below 6000~K with a 10~\kmps\ downdraft.  Figure 9b 
shows the emission for a model in which the gas above 8000~K is emitting with no downdraft and only 
the cooler gas with log(\nelect) $\geq$\ 8.6 is emitting with a 10~\kmps\ downdraft.  Both 
calculations give reasonable fluxes and electron densities averaged over the emitting gas (3.5 and 
5.5 $\times$10$^{8}$, respectively) high enough to begin to approximate the observed global values.
Thus calculations with the most realistic chromospheric model we can muster are at least 
{\it consistent} with Judge's suggestions about \C2].

This analysis has necessarily been speculative, although it is much better constrained by actual 
observations than any alternative analysis, e.g., one based on a traditional semi-empirical 
model, could have been.  I think it is fair to conclude that (1) the broad wings of \C2] likely 
come from formation in the lower wind (the broadening coming from differential expansion) and 
(2) C may not be ionized in the way expected from the standard non-LTE calculations of 
semi-empirical atmospheres but may be subject to some other source of ionization associated with 
the driving mechanism that picks out only part of the radial component of the turbulence 
distribution.  In all these calculations the wings of the \C2] line are broadened by formation 
in gas with significant expansion velocity projected into the line of sight.  The asymmetry 
favoring the blue wing over the red results from having some of the redshifted gas blocked by 
the star.  This effect is seen in actual stars, e.g., $\alpha$~Tau (Carpenter et al.\ 1991, 
Fig.\ 3) and $\gamma$~Cru (Carpenter et al.\ 1995, Fig.\ 2).  Furthermore, as a caveat, we should 
note (1) that C would be depleated by the first dredgeup (e.g., Luck \& Lambert 1985), although 
this effect is presumably included in my adopted C abundance, and (2) that the observations of 
$\zeta$~Aur binaries do not seem to show mostly neutral C at depth.  One should also keep in 
mind that \C2] may eventually destroy a basic assumption of this paper, and practically 
every other one written about chromospheres and winds of cool giants, that the chromosphere 
and wind are one continuous structure spread evenly over the face of the star.

\subsection{More Highly Ionized Species}

The emissions of highly ionized species such as \C3] and \C4\ do not fit readily into 
our model for 31~Cyg.  The level of ionization required is incompatable with the 
temperatures and densities measured in the outer winds of $\zeta$~Aur binaries as well as 
with the level of ionization actually measured in these winds.  However, the {\it clumping} 
of the gas throughout the chromosphere/wind gives us ample opportunity to incorporate more 
highly ionized material deep in the chromosphere and possible ways to excite it.  We have 
assumed that the gas is confined to blobs that fill $\sim$ 10\% of space, surrounded 
essentially by a vacuum filled with an, as yet unspecified, source of pressure.  The energy 
densities required to drive the wind through gradients of this pressure are certainly 
sufficiently high to allow higher ionization than the dominant levels we see in the shell lines.

\subsection{Velocity Structure from Fe II Self Reversals}

We may use our model for 31~Cyg to test the techniques used to deduce wind structure from 
shell lines in single stars.  The problem here is that such analyses for single stars give 
winds accelerating much more rapidly than those in $\zeta$~Aur binary components.  I have 
used my 31~Cyg model to predict what the velocities of the shell components of strong \Fe2\ 
lines in $\lambda$~Vel might be if it had the same velocity structure as 31~Cyg.  This is 
the star with the best evidence for wind acceleration in \Fe2\ self reversams, for which 
Carpenter et al.\ derived a wind with $\beta$=0.9, $v_{\infty}$=29--33 \kmps, and \Mdot = 
3$\times$10$^{-9}$ \MLrate.  In this simulation, I calculated the line-center radial optical 
depth into the atmosphere for the \Fe2\ and \Mg2\ lines measured by Carpenter et al.\ and 
determined the position and expansion velocity for $\tau$=2/3, as though the intrinsic line 
emission core comes from the center of the disc.  Table 3 gives the results, and Figure 10 
displays them.  These calculations assume the constraints of Table 2, with turbulence from 
Column 9.  For the mass-loss rate of 31~Cyg, the lines become optically thick very high in 
the wind, absent ionization of Fe$^{+}$ to Fe$^{+2}$ (Table 3, Column 5), but if we reduce 
the mass-loss rate somewhat to 9$\times$10$^{-9}$ \MLrate, to partially reflect Carpenter's 
lower value, and allow for ionization of Fe to Fe$^{+2}$ for \nelect\ $<$ 2$\times$10$^6$ 
(Table 2, Column 8), we get velocities in much better agreement with observed values, as 
seen in Figure 10.  Unfortunately we cannot apply this sort of analysis directly to the 
self-reversals of \Fe2\ lines in 31~Cyg because the emision is formed not in the inner 
chromosphere, as in single stars, but by scattering light from the B star in a huge shell 
surrounding the binary system.

These results show that we can actually reproduce the sort of observations taken to 
indicate a rapid acceleration of the chromosphere in a single star with a model 
accelerating much more slowly.  The secret for doing this is the realization that 
the observed absorption is limited by ionization of the outer wind as discussed in 
\S\ 4.2.  Therefore, there is really no compelling reason to believe the wind 
acceleration in the single windy giants is necessarily any faster than in binary 
components.  

There is evidence in occultations, as well as in the H$\alpha$ profiles, that 
chromospheres/winds of single stars are much more extended than commonly thought.
An occultation in H$\alpha$ emission for the single M supergiant 119 Tau at M2 Ib 
(White et al.\ 1982) and an image of $\alpha$~Ori at M2 Iab (Hebden et al.\ 1987) 
both showed H$\alpha$ emission, hence significant optical depths in the Lyman lines,
out to at least twice the photospheric radii.  Also, all of these cool supergiants 
show asymmetric H$\alpha$ profiles indicating the line core is formed in the wind 
at least some of the time (Mallik 1993).  Furthermore, in their model of the wind of 
$\alpha$~Ori, Harper et al.\ (2001) find a velocity/density structure giving even slower 
acceleration than we find in 31~Cyg or they found (Harper et al.\ 2005) in $\zeta$~Aur.

\section{DISCUSSION}

The importance of $\zeta$~Aur binaries is that they tell us just where the bulk of the gas 
is in the inner few stellar radii of the stars' chromospheres or winds.  We can measure 
column densities directly and get at least a reasonable idea of how dense and hot it is 
throughout this critical zone where most of the energy is injected into a star's wind.  
Furthermore, with models like those of \S 3.2, we know that this gas must be emitting 
the intrinsic lines that we would easily see absent the B companions.  The structure is 
therefore constrained in ways it simply cannot be in any analysis of a single star.  I have 
subjected the model for 31~Cyg, and by extension models for other $\zeta$~Aur binaries, to 
tests based on various sorts of emission from their chromospheres/winds.  That such a model 
reproduces the intrinsic \C2], \Al2], and thermal radio emission to $\sim$ 50\% with a 
minimum of fiddling is an excellent test of the reality of the structure derived.  
Furthermore, the clumping inferred for such models provides an excellent opportunity (n.b., 
this really is fiddling) to accomodate tenuous material emitting lines of the more highly 
ionized species.  This agreement is reassuring, inasmuch as most of the properties of the 
model are constrained by {\it observations}.  The chief uncertainties involve ionization of 
H throughout the atmosphere, which is {\it not} measured well observationally, and ionization 
of H and the metals in the outer wind (\Telect\ $>$ 9,000K) where many of the self reversals 
of strong lines would probably be formed.

\subsection{Alfven-Wave Models}

There are several fairly extensive investigations of the conditions required for driving winds 
by Alfven waves that provide potentially testable predictions (Hartmann \& McGregor 1980; 
Hartmann \& Avrett 1981; Holzen, Fl\aa, \& Leer 1983; Kuin \& Ahmad 1989).  For the most part 
these suffer from limitations of assuming the gas is effectively coupled to the fields (i.e., 
fully ionized), from not exploring the mechanisms of transfer of momentum from wave to gas 
realistically (although Holzer et al.\ did begin this process), and therefore from relying on
somewhat nonphysical theories of how the waves are damped.  

There are a couple of ways of applying a wave model for driving a wind.  In the more fundamental 
approach, one would use theories of how such waves are generated and how they interact with partially 
ionized media to predict the structure and properties of winds theoretically.  Holzer et al.\ 
(1983, \S\ VI) actually calculated some models of this sort.  Unfortunately, the theories required 
to do that are difficult to apply, and a knowledge of magnetic fields 
of actual stars that would support such waves is lacking.  The alternative is to form a semi-empirical 
model for wave-driven winds, somewhat like the chromospheric models I have been discussing, and determine 
what the properites it must have to fit the observed structure of a wind (viz., mass-loss rate, 
terminal velocity, velocity-density structure).  A good example of this second approach is the analyses 
of $\zeta$~Aur binaries by Kuin \& Ahmad.  Their models find that damping of the wave amplitude must 
decrease with height to fit observed velocity profiles, as one might well expect theoretically to keep 
the terminal velocity consistent with observation.  Their models also give predictions of the level 
of turbulence in the chromosphere/wind by associating the lateral displacements of gas by such waves 
with the observed Doppler widths of gas in the atmospheres of these stars.  Predicted Doppler widths, 
both by Kuin \& Ahmad  and Hartmann \& Avrett seem to be larger than the observed turbulence, both 
in shell absorptions in $\zeta$~Aur binaries and in the optically-thin emission lines of single windy 
giants.  Furthermore, since Alfven waves are transverse, the ``turbulence" would be anisotropic 
to first order.  This prediction is at odds with the observation of isotropic turbulence, although 
it is based on the idealization of radial magnetic fields, while the actual topology might well be 
more complicated (see, e.g., Cassinelli et al.\ 1995).

Kuin \& Ahmad found from their semi-empirical models that the damping length for Alfven waves 
must increase with height.  The most convincing mechanism for transferring energy from the wave 
to the gas, i.e., damping it, is so-called ion-neutral friction in which there is a phase lag 
between the wave's transferring momentum to ions and the ions' subsequently transferring it to 
neutrals.  Hartmann \& McGregor discussed this mechanism, although they had no way of applying 
it {\it a priori}.  For waves with long periods, the transfer can be so rapid that the neutrals 
are effectively bound to the ions through elastic collisions, and there is little dissipation 
of wave energy or transfer of momentum.  For high frequency, on the other hand, the neutrals 
cannot respond fast enough to the passage of a wave to partake in its displacements, and the ions 
just stir them up and dissipate wave motion as heat.  Since the dissipated wave energy would go 
primarily into heating, this mechanism would be better for heating the chromosphere/wind than 
driving its outflow.  Most of the wave energy would be available for heating the wind but not 
for driving it, since the measured temperatures in these winds are much too low to drive them
by thermal expansion (cf. \S\ 3).  Presumably the {\it momentum} of the wave would be 
transferred into wind motion.  However, since the ratio of energy to momentum ($E/p$) goes as 
1/$v$, with the Alfven speed generally larger than either a thermal or turbulent velocity, 
philosophically an attractive driving mechanism would transfer its energy to heat or some other 
mass motion as an intermediate stage.  This is why it is so hard to drive winds with radiation 
pressure ($E/p$ $\sim$ 1/$c$).  If all the waves had the same frequency, we would expect the 
momentum to be deposited in a narrow range of density, hence height, contrary to measurements 
of the acceleration of actual winds.

The mechanism of ion-neutral friction would imply a damping length that decreased with height 
contrary to Kuin \& Ahmad's result.  However, it also implies that waves of different frequencies 
would be absorbed at different height and, therefore, that the {\it spectrum} of Alfven waves would 
determine the velocity profile of the wind.  At this point our knowledge of the photospheric 
motions that might be exciting Alfven waves and the spectrum they would produce seems too 
sketchy to make any testable predictions about a wind's velocity structure.

\subsection{An Alternative Wind Mechanism}

Let us now take the liberty to speculate about a different way of driving the mass loss of these 
windy giants.  We have developed here a picture of what conditions are required to drive the wind 
of one particularly well observed wind structure.  Pressures required are an order of magnitude 
greater than those of the implied density/temperature structure.  That the microturbulence 
required to fit line shapes and widths seems to be rather isotropic means that we likely are 
not simply seeing the effects of globally organized Alfven waves passing through the gas, as 
proposed by Hartmann \& McGregor (1980) and implied by models of Airapetian et al.\ (2000), 
for example.  There may be another way of supporting a wind with magentic fields, namely using 
{\it chaotic fields} emerging from the star and diffusing through the gas into space to drag the 
gas along with it and away from the star.  Some form of this idea was implicit in our previous 
musings about the wind structure of 31~Cyg (Eaton \& Bell 1994, \S\ 6), and Mullan et al.\ 
(1998) may have waved at it in passing.  This very speculative picture is fundamentally different 
from the Alfven model in that the magnetic field being lost can constitute a moderate amount of 
luminosity.  In the standard Alfven model, gas is driven $\sim$ radially from the star along 
magnetic flux lines anchored permanently in the stellar surface.  Here, magnetic flux would be 
lost at $\sim$ the same rate as the gas and constitute a significant component of the energy loss 
in the wind through its adiabatic expansion.  Such chaotic field would give a much more isotropic 
pressure, which would impress itself on the random velocities of the gas.  We may estimate the effect 
by assuming the pressures driving the wind are in equipartition with the kinetic energy of random 
motions (turbulence) in the gas.  For this condition the driving energy is \ensuremath{\onehalf{NkT}} 
per degree of freedom (three of them), $N$ being the number of particles in a random blob of mass $M$.
The kinetic energy of the blob is \ensuremath{\onehalf{Mv_{\rm equ}^2}} (per degree of freedom), so 
that equipartition gives \ensuremath{Mv_{\rm equ}^2} = $NkT$, or
\begin{equation}
    v_{\rm equ} =  (k/<m>)^{0.5} T_{\rm therm}^{0.5} = 0.08  T_{\rm therm}^{0.5}  \\
\end{equation}  
where we have taken the average mass per particle to be 1.3$m_{\rm H}$.  This is to within a factor 
of $\surd\gamma$ of the sound velocity, but for our elevated artificial temperature.  For values 
of the driving temperature, $T_{\rm therm}$ (Table 2, Column 4), we get the turbulent velocities given 
in Table 2, Column 9.  These values are comparable to the line-of-sight random velocities measured in cool 
(super)giants, and this fact argues that the driving force must be able to produce the random motions 
observed.  It is unlikely, therefore, to be global Alfven waves.  Of course, the pressure of these 
turbulent motions is itself a major source of momentum and energy in extending the atmosphere and 
driving the wind.  

Energy input determines the magnetic fields required in this model, since the magnetic energy density 
must be greater than the energy per unit volume required to lift the mass out of the potential 
of the star, B$^2$/8$\pi$ $>$ G$M_{\rm K}$$\rho$/\Ro.  For our model, this leads to a field strength 
of 25 Gauss.  If we assume the energy loss is double the potential energy from kinetic energy of 
the wind and emission from it, the field strength increases only to 35 Gauss.

This kind of driving has the advantage over Alfven waves of being able to begin to explain the 
variation of mass loss from star to star in a way related to stellar structure and evolution.  
In our simulation of driving with thermal profile, the expansion-velocity structure is determined 
by the temperature (i.e., energy-density) profile, 
while the mass-loss rate is arbitrary, determined by supplying enough energy at some \Mdot\ to 
maintain the energy-density profile.  With magnetic-flux emergence as the driving mechanism, the 
magnetic energy corresponds to the infusion of thermal energy in the coronal model; mass loss 
rate, therefore, is proportional to the rate at which magnetic field emerges.  This is very 
attractive in that there are indications that the winds of the giants vary in response to changes 
that can re-excite dynamos in their cores.  See Mullan \& MacDonald (2003) for changes 
of mundane giants, and recall the shells episodically thrown off by pulsing AGB stars.  Alfven 
waves, on the other hand, would most likely be excited by convection, therefore be proportional to 
luminosity, and be little affected by the strength of the passive magnetic fields serving as 
their medium.

As an alternative to chaotic magnetic fields filling the voids in chromospheric gas, we may imagine 
Alfven waves trapped in the cavities between ionized blobs.  These waves would have speeds approaching 
the speed of light as the density dropped, and would be reflected off the blobs if their frequencies 
were below the cyclotron frequency.  The critical frequency would rise as the material became more 
highly compressed (denser).  Eaton \& Bell actually had this mechanism in mind as the driving pressure 
in such atmospheres.  There must be a rich optics of these Alfven waves waiting to be discovered.  

\section{SUMMARY}

I have constructed a model for the chromosphere and wind of 31~Cyg which is based on measurements 
of physical properties in the outer atmospheres of 31~Cug and other classical $\zeta$~Aur binaries.
It goes beyond other such models in that it derives the poorly understood turbulence and clumping 
of the gas from the pressures driving the wind's expansion.  It predicts emission of optically-thin 
lines and microwave continuum to within 50~\%\ of observed values, excellent agreement in the 
circumstances.  In this model, the momentum flux required to drive the wind determines the 
stratification of the chromosphere where intrinsic lines would be formed.  That momentum flux 
gives much lower densities than the stratification from thermal momentum flux alone, with the 
consequence that the gas must be clumped to produce the observed flux in intrinsic lines.  We 
also find that this model can reproduce most of the properties of single stars' chromospheric 
spectra and argue that the evidence for fundamental differences between single stars and these 
binary components is rather weak.

We must keep in mind, however, that there are some inconsistencies in this picture.  The 
ionization of C, for instance, is a problem.  It cannot be mostly singly ionized, as it 
seems to be observationally, without giving fluxes much larger than observed, and the model 
for 31~Cyg does not predict the redshifts seen in single supergiants.  \C2] multiplet ratios 
from the model likewise would not predict the large global electron densities found from 
\C2] in single stars.  However, there are also inconsistencies in the interpretation of \C2] 
in the single stars themselves, as I discuss in \S\ 4.4.

\acknowledgements
I dedicate this paper to John S.\ Mathis who seemed to have the visceral intuitive grasp of physics 
required to address the sort of problems I have discused in this paper.  If I'd had the sense 
to take his advice to work in this area when I was in graduate school, I probably would have 
written it years ago.  This research has been supported by NSF through grant HRD-9706268 and 
NASA through grant NCC5-511 to TSU.

\clearpage

%%Table 1--Line Strengths

\begin{deluxetable}{cccccccc}
\tabletypesize{\scriptsize}
\tablecaption{Line Strengths in Cool Giants and Models}
\tablewidth{0pt}
\tablehead{
\colhead{Star} & \colhead{Spectral} & \colhead{$V$} & 
\colhead{f(Ly$\alpha$)} & \colhead{f(2325)} & \colhead{f(2669)} & 
\colhead{EW(H$\alpha$)} & \colhead{log(\nelect)}\\ 
\colhead{} & \colhead{Type} & \colhead{$(B-V)$} & 
\colhead{f$_{line}$/f$_V$} & \colhead{f$_{line}$/f$_V$} & 
\colhead{f$_{line}$/f$_V$} & \colhead{f$_{line}$/f$_V$}& } 
\startdata
$\alpha$ Boo & K1 III & $-$0.04 & $\geq$1.4$\times$10$^{3}$ & 35.7                & 83.1                & 1.12 &9.7\\
             &        &    1.23 & $\geq$3.7$\times$10$^{-2}$& 9.4$\times$10$^{-4}$& 2.2$\times$10$^{-3}$&      &   \\ 
$\alpha$ Tau & K4 III &    0.85 & $\geq$340                 & 31.3                & 30.7                & 1.12 &9.0\\
             &        &    1.44 & $\geq$2,0$\times$10$^{-2}$& 1.9$\times$10$^{-3}$& 1.88$\times$10$^{-3}$&     &   \\ 
$\beta$ Gru  & M5 III &    2.13 &   $>$50                   & 31.7                & 28.8             & \nodata &8.5\\
             &        &    1.57 & $>$9,7$\times$10$^{-3}$   & 6.2$\times$10$^{-3}$& 5.6$\times$10$^{-3}$&      &   \\ 
$\lambda$ Vel& K5 Ib-II&   2.21 &    \nodata                &  \nodata            &  12                 & 1.52 &8.9\\
             &        &         &                           &                     & 2.5$\times$10$^{-3}$&      &   \\ 
31 Cyg       & K4 I   &    3.79 &    \nodata                & $<$5                &  2.3                & 1.50 &\nodata\\
             &        &    1.28 &                           &$<$5$\times$10$^{-3}$& 2.1$\times$10$^{-3}$&      &   \\ 
32 Cyg       & K4 I   &    3.98 &    \nodata                &$\leq$5$\pm$100\%    &  3.4                & 1.75 &\nodata\\
             &        &    1.52 &                      &$\leq$5.3$\times$10$^{-3}$& 3.6$\times$10$^{-3}$&      &   \\ 
$\zeta$ Aur  & K4 Ib  &    3.79 &    \nodata                &$\leq$4$\pm$100\%    &  2.2                & 1.57 &\nodata\\
             &        &    1.22 &                      &$\leq$3.7$\times$10$^{-3}$& 2.0$\times$10$^{-3}$&      &   \\ 
Model 1      &        &         &    \nodata                & 56                  & 28                & \nodata &9.8\\
(3\% ioniz)  &        &         &                           &                     &                     &      &   \\ 
Model 2      &        &         &    \nodata                & 9.1                 & 3.6               & \nodata &8.5\\
(var $<$10\%)&        &         &                           &                     &                     &      &   \\ 
Model 3      &        &         &    \nodata                & 6.7                 & 3.0               & \nodata &8.6\\
(var $<$3\%) &        &         &                           &                     &                     &      &   \\ 
\enddata
\tablecomments{~Line fluxes at the Earth are in 10$^{-13}$ ergs cm$^{-2}$s$^{-1}$. 
f(line)/f$_V$ is in \AA\ and assumes f$_V$ = 3.65$\times$10$^{-9}$ 
ergs~cm$^{-2}$s$^{-1}$\AA$^{-1}$ at $V$ = 0.0.  Values of log(\nelect) in Col.\ 8 
come from C~II] multiplet ratios as noted in the text.}
\end{deluxetable}

\clearpage

%%Table 2--Model for 31 Cyg

\begin{deluxetable}{crrcrcccc}
\tabletypesize{\tiny}
\tablecaption{Details of Model for 31 Cyg}
\tablewidth{0pt}
\tablehead{
\colhead{R} & \colhead{$v_{\rm exp}$}   & \colhead{$T_{\rm exc}$}   & \colhead{$T_{\rm therm}$} & 
\colhead{CF}& \colhead{log($n_{\rm H}$)}& \colhead{log(\nelect)}    & \colhead{log(\nelect)}    & 
\colhead{$v_{\rm equ}$} \\
\colhead{} & \colhead{} & \colhead{} & \colhead{} & \colhead{} & \colhead{} & 
\colhead{(3\%)} & \colhead{(var)\tablenotemark{a}} & \colhead{} \\
\colhead{(1)} & \colhead{(2)} & \colhead{(3)} & \colhead{(4)} & \colhead{(5)} & \colhead{(6)} & 
\colhead{(7)} & \colhead{(8)} & \colhead{(9)}
}
\startdata
 7000.0 & 83.8 &  12500. &  20000. &  1.60 &  4.53 &  3.21 &  3.73 & 11.3 \\
 6200.5 & 83.0 &  12500. &  20000. &  1.60 &  4.64 &  3.32 &  3.84 & 11.3 \\
 5466.3 & 82.1 &  12500. &  20000. &  1.60 &  4.75 &  3.43 &  3.95 & 11.3 \\
 4794.5 & 81.0 &  12500. &  20000. &  1.60 &  4.87 &  3.55 &  4.07 & 11.3 \\
 3897.8 & 79.1 &  12500. &  20000. &  1.60 &  5.06 &  3.74 &  4.26 & 11.3 \\
 3369.9 & 77.4 &  12500. &  24468. &  1.96 &  5.20 &  3.96 &  4.49 & 12.5 \\
 3126.0 & 76.5 &  12500. &  30660. &  2.45 &  5.27 &  4.13 &  4.66 & 14.0 \\
 2894.9 & 75.5 &  12500. &  36526. &  2.92 &  5.34 &  4.28 &  4.80 & 15.2 \\
 2676.2 & 74.3 &  12500. &  42075. &  3.37 &  5.41 &  4.42 &  4.94 & 16.3 \\
 2469.7 & 73.1 &  12500. &  47316. &  3.79 &  5.49 &  4.55 &  5.07 & 17.3 \\
 2275.0 & 71.8 &  12500. &  52258. &  4.18 &  5.57 &  4.67 &  5.19 & 18.2 \\
 2091.8 & 70.3 &  12500. &  56909. &  4.55 &  5.65 &  4.79 &  5.31 & 19.0 \\
 1919.7 & 68.7 &  12500. &  61278. &  4.90 &  5.74 &  4.90 &  5.43 & 19.7 \\
 1758.3 & 66.9 &  12500. &  65374. &  5.23 &  5.82 &  5.02 &  5.54 & 20.4 \\
 1607.3 & 64.9 &  12500. &  69206. &  5.54 &  5.92 &  5.14 &  5.66 & 21.0 \\
 1466.4 & 62.8 &  12500. &  72783. &  5.82 &  6.01 &  5.25 &  5.77 & 21.5 \\
 1335.2 & 60.4 &  12500. &  76113. &  6.09 &  6.11 &  5.37 &  5.89 & 22.0 \\
 1213.3 & 57.8 &  12500. &  79205. &  6.34 &  6.21 &  5.49 &  6.01 & 22.4 \\
 1100.5 & 55.0 &  12306. &  82068. &  6.67 &  6.32 &  5.62 &  6.14 & 22.8 \\
  996.4 & 51.9 &  11902. &  84712. &  7.12 &  6.43 &  5.76 &  6.28 & 23.2 \\
  900.6 & 48.5 &  11508. &  87143. &  7.57 &  6.54 &  5.90 &  6.42 & 23.5 \\
  812.7 & 45.0 &  11103. &  89373. &  8.05 &  6.67 &  6.05 &  6.57 & 23.8 \\
  732.5 & 41.1 &  10709. &  91408. &  8.54 &  6.80 &  6.20 &  6.73 & 24.1 \\
  659.6 & 37.1 &  10308. &  93259. &  9.05 &  6.93 &  6.37 &  6.89 & 24.3 \\
  593.6 & 32.8 &   9910. &  94934. &  9.58 &  7.08 &  6.53 &  7.06 & 24.6 \\
  534.2 & 28.5 &   9515. &  95000. &  9.98 &  7.23 &  6.71 &  7.23 & 24.6 \\
  481.1 & 24.1 &   9133. &  95000. & 10.40 &  7.39 &  6.89 &  7.41 & 24.6 \\
  433.8 & 19.8 &   8755. &  95000. & 10.85 &  7.57 &  7.08 &  7.60 & 24.6 \\
  392.2 & 15.7 &   8388. &  95000. & 11.33 &  7.76 &  7.29 &  7.81 & 24.6 \\
  355.7 & 12.0 &   7193. &  95000. & 13.21 &  7.96 &  7.56 &  7.78 & 24.6 \\
 324.04 & 8.66 &   6830. &  83520. & 12.23 &  8.18 &  7.75 &  7.76 & 23.0 \\
 296.92 & 5.92 &   6492. &  69960. & 10.78 &  8.42 &  7.93 &  7.74 & 21.1 \\
 273.96 & 3.77 &   6178. &  58482. &  9.47 &  8.69 &  8.14 &  7.77 & 19.3 \\
 254.82 & 2.36 &   5883. &  48912. &  8.31 &  8.95 &  8.35 &  7.81 & 17.6 \\
 239.15 & 1.44 &   5625. &  41077. &  7.30 &  9.23 &  8.57 &  7.87 & 16.2 \\
 226.61 & 0.81 &   5410. &  34803. &  6.43 &  9.52 &  8.80 &  7.98 & 14.9 \\
 216.83 & 0.42 &   5244. &  29917. &  5.70 &  9.84 &  9.08 &  8.16 & 13.8 \\
 209.49 & 0.20 &   5105. &  26245. &  5.14 & 10.20 &  9.39 &  8.39 & 12.9 \\
 204.23 & 0.11 &   4998. &  23614. &  4.72 & 10.48 &  9.63 &  8.57 & 12.2 \\
 200.70 & 0.06 &   4930. &  21850. &  4.43 & 10.75 &  9.87 &  8.77 & 11.8 \\
 198.56 & 0.04 &   4905. &  20781. &  4.24 & 10.94 & 10.04 &  8.92 & 11.5 \\
 197.46 & 0.04 &   4905. &  20231. &  4.12 & 10.98 & 10.07 &  8.95 & 11.3 \\
 197.06 & 0.04 &   4905. &  20029. &  4.08 & 10.98 & 10.07 &  8.95 & 11.3 \\
 197.00 & 0.04 &   4905. &  20000. &  4.08 & 10.98 & 10.07 &  8.95 & 11.3 \\
\enddata
\tablenotetext{a}{Ionization as in Figure 5 but limted to 10\% in 
the outer chromosphere.}
%% Note to Table 2: For 31 Cyg, we assume D=473 pc; R = 197 \Rsun, M$_K$ = 11.7 \Msun, M$_B$ = 7.1 \Msun, etc.
%% Mdot = 3.0E-8 & \Msun~s$^{-1}$
\end{deluxetable}

\vfill

\clearpage

%%Table 3--Expansion Velocties for Shell Lines

\begin{deluxetable}{lcccccc}
\tabletypesize{\scriptsize}
\tablecaption{Calculated Expansion Velocties for Shell Lines}
\tablewidth{0pt}
\tablehead{
\colhead{Multiplet} & \colhead{$\lambda$} & \colhead{log(LSF)} & 
\colhead{$v_{\lambda Vel}^{\rm obs}$} & \colhead{$v_{\rm 31 Cyg}^{\rm calc}$} & 
\colhead{$v_{\lambda Vel}^{\rm calc}$}\\
\colhead{} & \colhead{(\AA)} & \colhead{@6000K} & \colhead{(\kmps)} & \colhead{(\kmps)} & 
\colhead{(\kmps)}} 
\startdata
Fe II UV1    & 2599.40&    0.86& $-$35.0& $-$83.4& $-$38.0 \\
Mg II UV1    & 2795.52&    0.66& $-$31.6& $-$83.8& $-$38.0 \\
Mg II UV1    & 2802.70&    0.36& $-$33.1& $-$83.7& $-$38.0 \\
Fe II UV1    & 2598.37&    0.35& $-$33.4& $-$82.4& $-$34.7 \\
Fe II UV1    & 2585.88&    0.30& $-$34.7& $-$82.2& $-$33.9 \\
Fe II UV1    & 2607.09&    0.25& $-$34.1& $-$82.0& $-$33.9 \\
Fe II UV3    & 2332.80&    0.19& $-$32.2& $-$81.8& $-$33.0 \\
Fe II UV62   & 2755.73&    0.04& $-$32.0& $-$81.0& $-$34.7 \\
Fe II UV3    & 2364.83& $-$0.01& $-$30.0& $-$80.6& $-$30.5 \\
Fe II UV1    & 2625.66& $-$0.05& $-$33.7& $-$80.3& $-$30.1 \\
Fe II UV32   & 2739.55& $-$0.05& $-$27.2& $-$80.3& $-$33.9 \\
Fe II UV3    & 2338.01& $-$0.06& $-$32.0& $-$80.3& $-$30.5 \\
Fe II UV1    & 2617.62& $-$0.16& $-$33.4& $-$79.4& $-$28.4 \\
Fe II UV35   & 2362.02& $-$0.62& $-$17.8& $-$69.8& $-$20.3 \\
Fe II UV35   & 2331.30& $-$0.75& $-$29.0& $-$64.0& $-$16.6 \\
Fe II UV35   & 2366.59& $-$0.84& $-$13.5& $-$64.0& $-$15.4 \\
Fe II UV64   & 2593.72& $-$0.91& $-$34.0& $-$59.2& $-$19.4 \\
Fe II UV64   & 2591.52& $-$0.93& $-$16.5& $-$53.9& $-$18.2 \\
Fe II UV32   & 2736.97& $-$0.99& $-$16.4& $-$50.1& $-$17.0 \\
Fe II UV35   & 2354.89& $-$0.99& $-$15.9& $-$49.8& $-$11.8 \\
Fe II UV63   & 2761.81& $-$1.37&  $-$7.9& $-$25.8&  $-$7.4 \\
Fe II UV63   & 2772.72& $-$1.48&($-$0.8)& $-$19.8&  $-$5.4 \\
Fe II UV260  & 2741.40& $-$2.86&($-$2.5)&  $-$0.5&     0.0 \\
Fe II UV32   & 2732.41& $-$3.11&    +2.7&  $-$0.2&     0.0 \\
Fe II UV32   & 2759.34& $-$3.18&($-$2.4)&  $-$0.1&     0.0 \\
\enddata
\end{deluxetable}

\clearpage

%%Figure 1--Coronal wind model
%%This figure comes from winds/src/31cygmod-fig01.ps

\begin{figure}
\epsscale{0.65}
\plotone{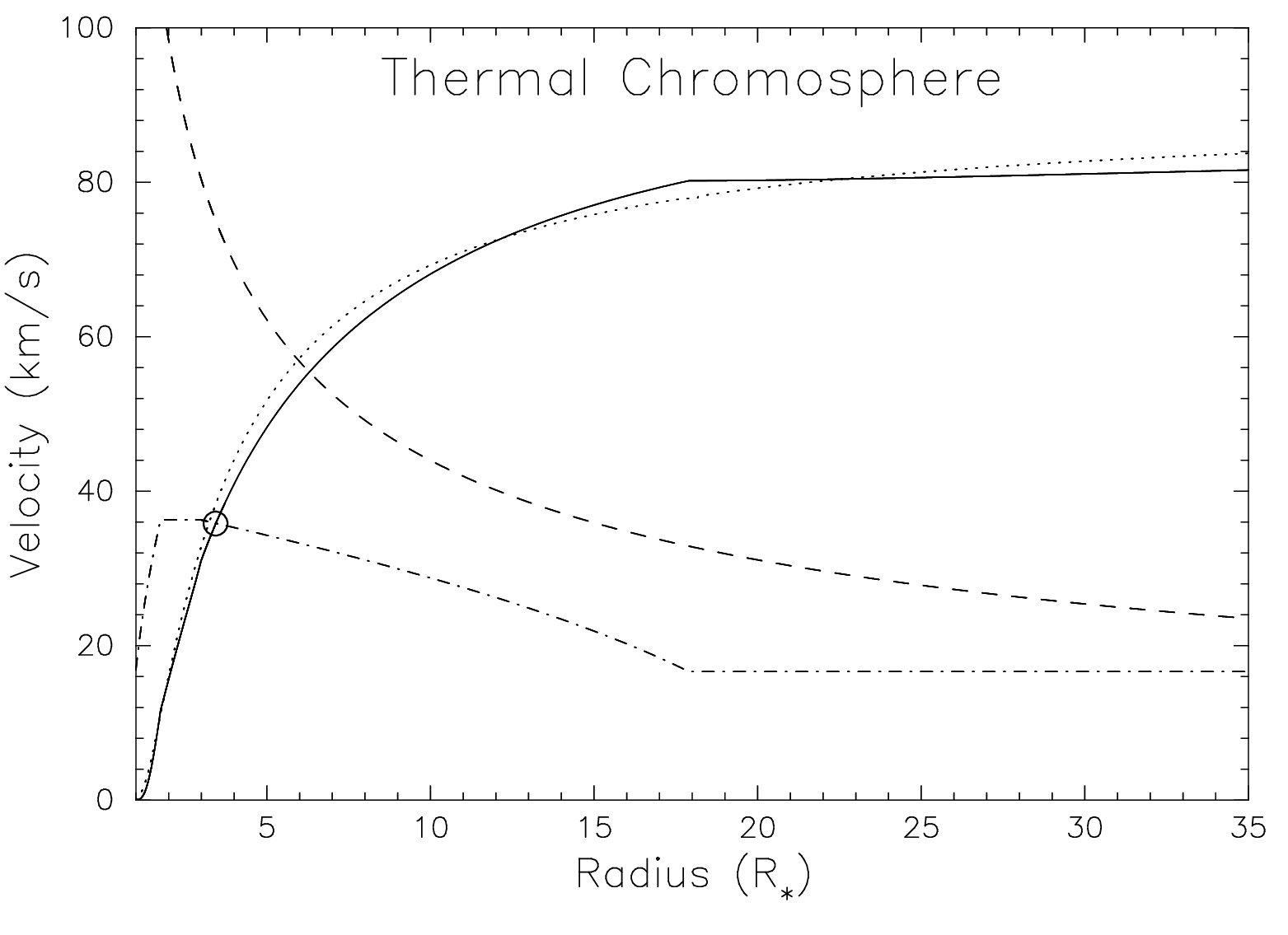}
\caption{Coronal wind model to fit the derived velocity structure of 31~Cyg.
The dotted line is the velocity profile derived for 31~Cyg by Eaton \& Bell (1994), 
while the solid curve is the velocity profile calculated with the temperature 
profile of Equations 2.  Other curves are the local escape velocity (dashed) 
and sound speed for the velocities in Equations 2 (dot-dashed).  The circle 
shows the sonic point in this model.\label{fig1}}
\end{figure}

%%Figure 2--Inferred clumping
%%This figure comes from winds/analysis/clumping2.ps

\begin{figure}
\epsscale{0.65}
\plotone{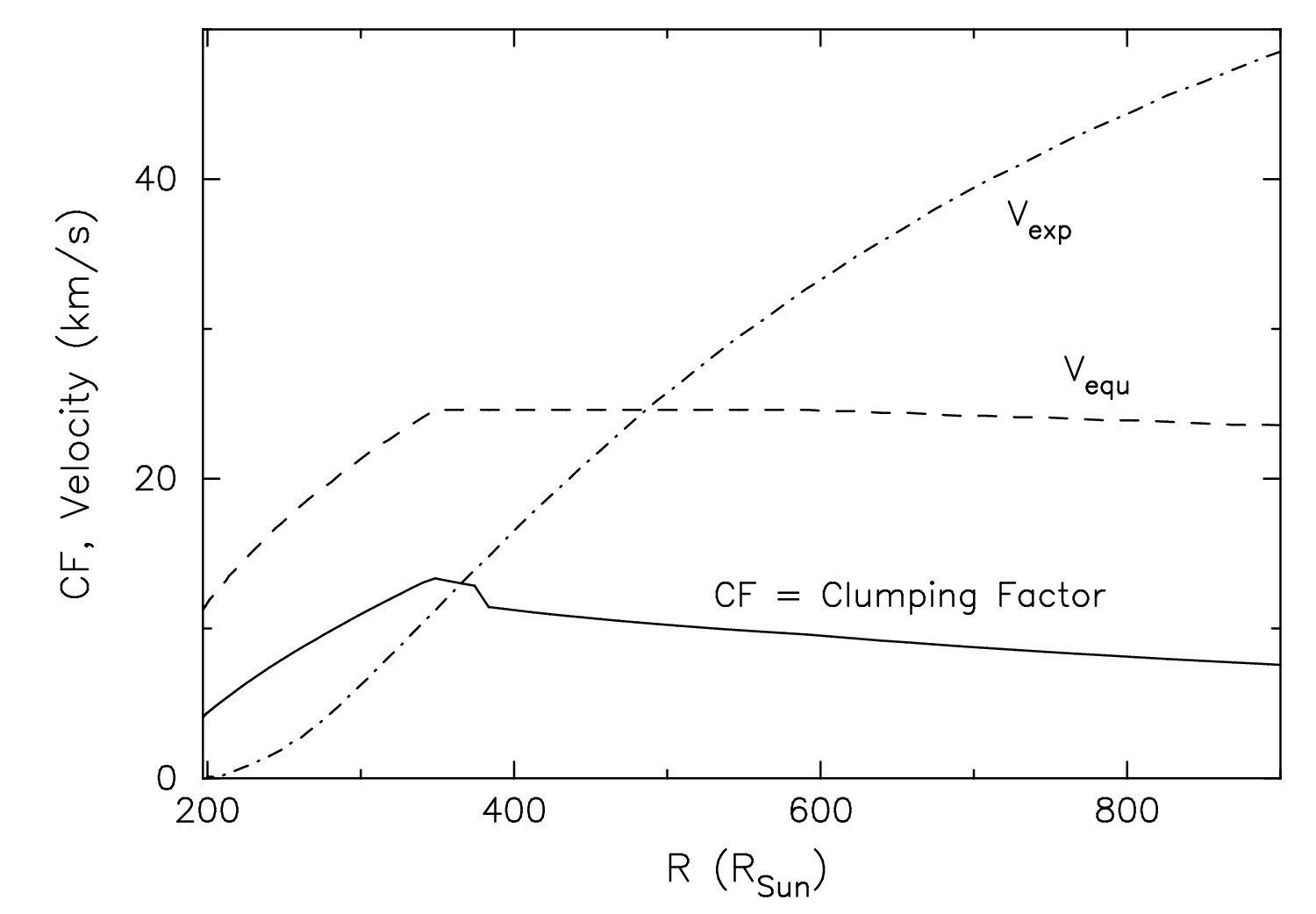}
\caption{Inferred clumping for the 31~Cyg model.  Solid curve is the calculated 
clumping factor, CF, and the dashed line is the equipartition velocity defined 
by Equation 7.  The atmospheric expansion velocity, $v_{\rm exp}$ is shown for comparison.
\label{fig2}}
\end{figure}

%%Figure 3--Profile for optically-thin lines
%%The panels in this figures here come from winds/analysis/31cygmod-fig03a.ps and
%%winds/analysis/31cygmod-fig03b.ps

\begin{figure}
\begin{center}
\epsscale{1.05}
\plottwo{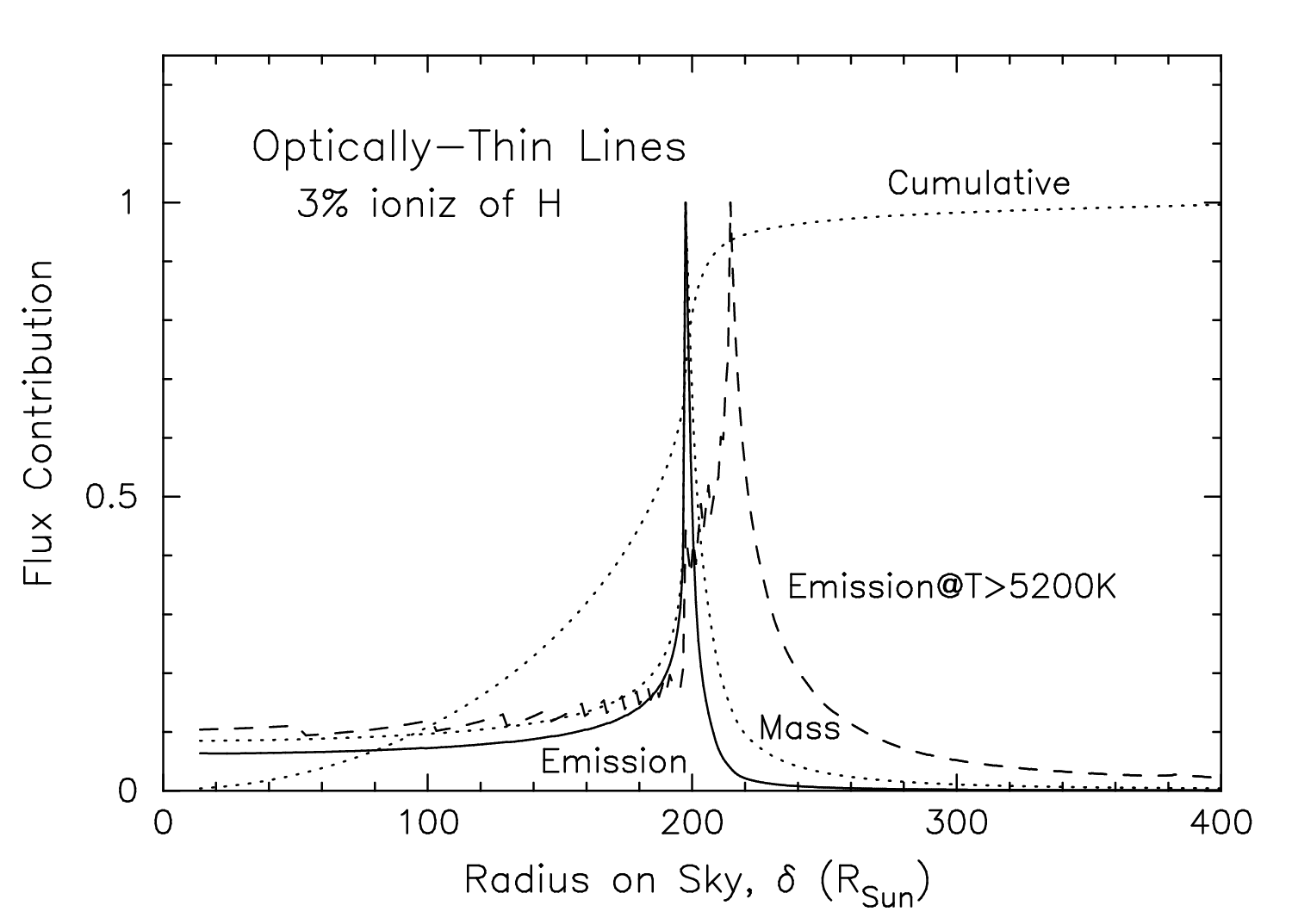}{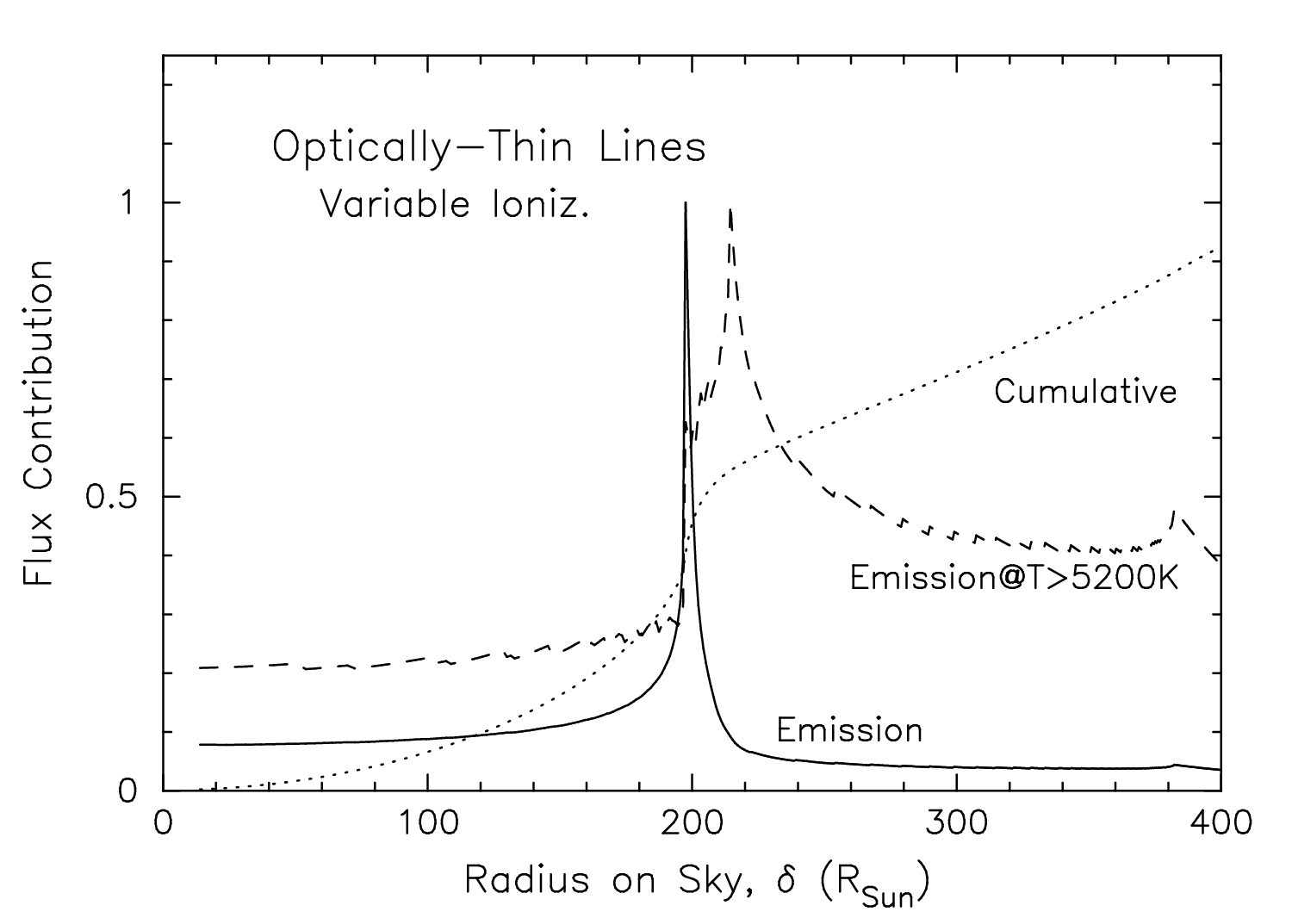}
\end{center}
\caption{Calculated intensity of emission over the stellar disc for optically-thin 
emission lines. (left) This figure assumes the emitter is in the same ionization stage  
throughout the chromosphere, \nelect\ in Column 7 of Table2.  (right) This figure is for 
the more realistic level of ionization (\nelect\ in Column 8 of Table2).
\label{fig3}}
\end{figure}

%%Figure 4--Profile of RADIO emission
%%This figure comes from winds/analysis/radio.radio-emis2.ps

\begin{figure}
\epsscale{0.75}
\plotone{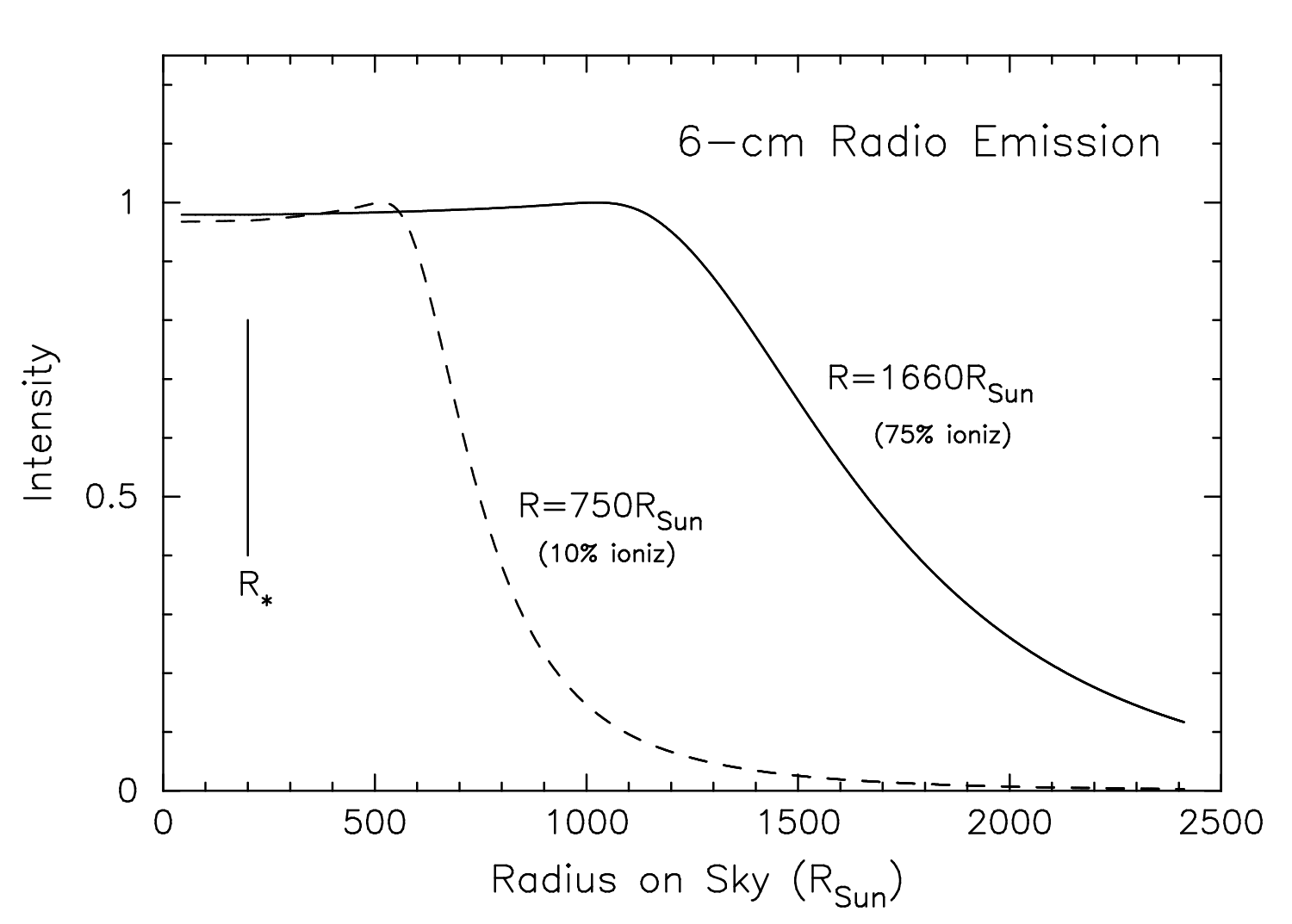}
\caption{Calculated intensity of 6.17-cm radio emission over the stellar disc for two 
models: dashed curve for the model with variable ionization limited to 10\% (\nelect\ in 
Column 8 of Table 2) and solid curve for the same model with 75\% ionization of H 
in the outer atmosphere.  This latter distribution would project a disc of 38 mas and 
have a spectral index $\alpha$ = 1.08, comparable to the few values measured by 
Drake \& Linsky (1986).
\label{fig4}}
\end{figure}

%%Figure 5--Ionization from PANDORA
%%This figure comes from winds/pandora/nenh-raw.ps

\begin{figure}
\epsscale{0.55}
\plotone{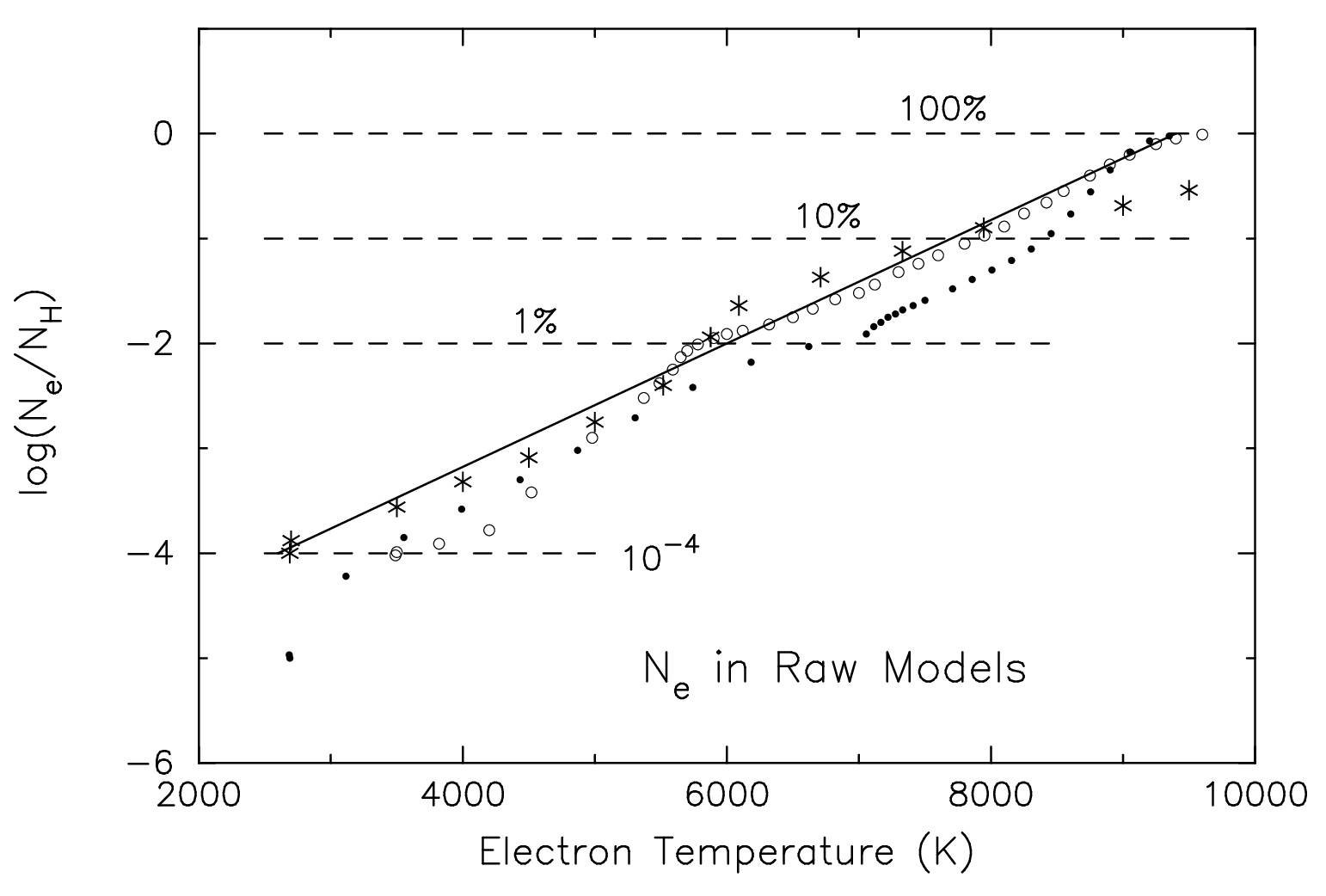}
\caption{Calculated ionization in three semi-empirical models for cool giant stars.  
The graph gives the ratio of electron density to total H density as a function of 
temperature.  At low temperature, where ionization of H is low, the electron density 
is dominated by metals.  At high temperature H becomes completely ionized, in these 
models, if not in actual stars.  The region in which most of the chromospheric lines 
are emitted has a H ionization of a few percent.  The solid curve shows the ralation 
I have adopted to relate electron density to excitation temperature in the calculations 
with variable ionization of H.  Discrete symbols represent calculations for three models:
dots, $\alpha$~Tau (Eaton 1995: Table 4, Model t8), circles, $\epsilon$~Gem, and
asterisks, $\zeta$~Aur.
\label{fig5}}
\end{figure}

%%  at_t8.top   (dots), eps_prd.top (circles), and za_t2.top   (asterisks * 3)

%%Figure 6--Excitation Temperatures
%%This figure comes from winds/iue/texc-rhox.ps

\begin{figure}
\epsscale{0.55}
\plotone{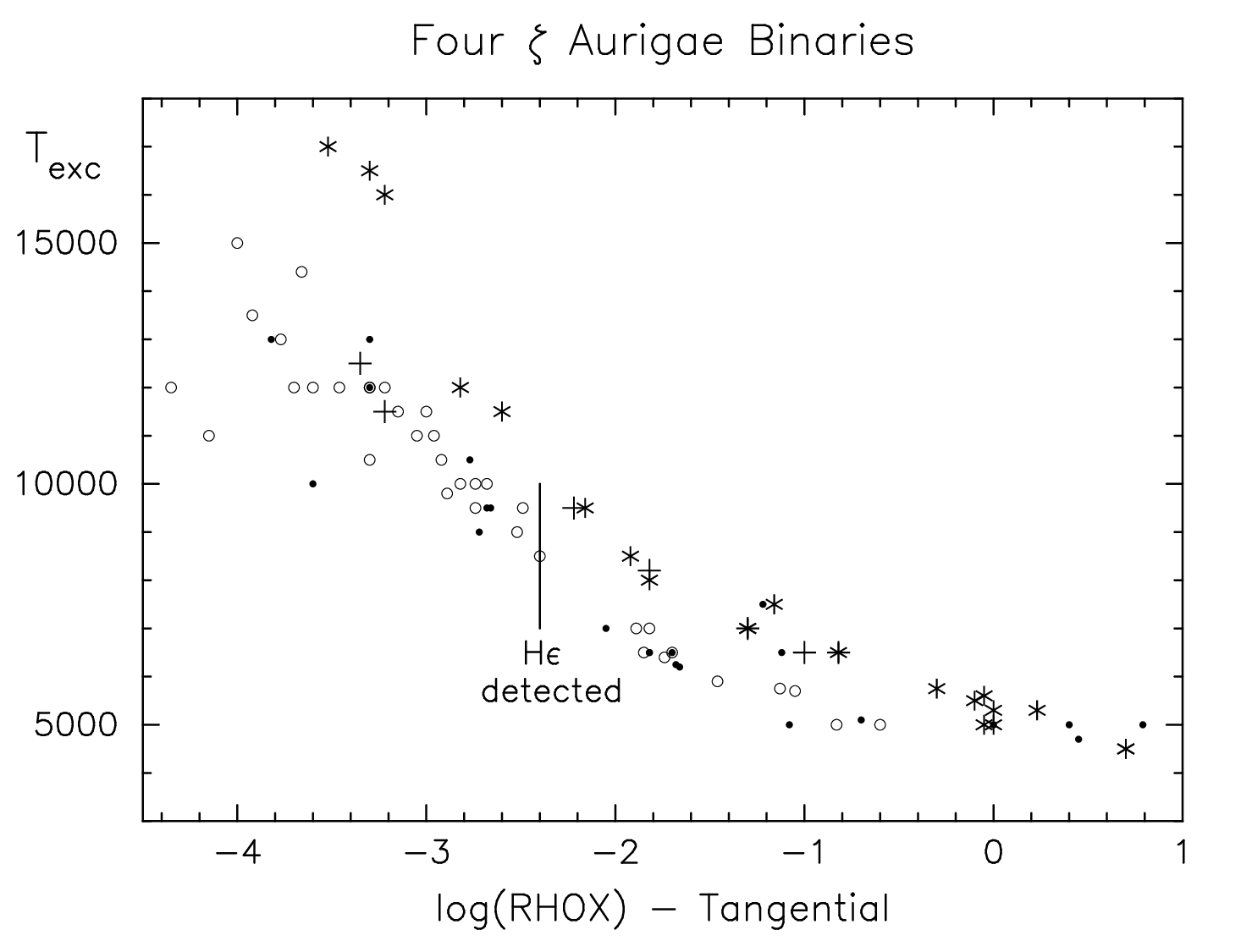}
\caption{Excitation temperature, \Texc,  vs.\ tangential mass column density, \Rhox, 
as measured in four $\zeta$~Aur systems.  Plusses are for $\zeta$~Aur, asterisks 
for 32~Cyg,  circles for 31~Cyg, and dots for 22~Vul.
\label{fig6}}
\end{figure}

%%Figure 7--Profiles for two ions
%%The two panels in this figures here come from winds/analysis/uniform

\begin{figure}
\begin{center}
\epsscale{0.85}
\plotone{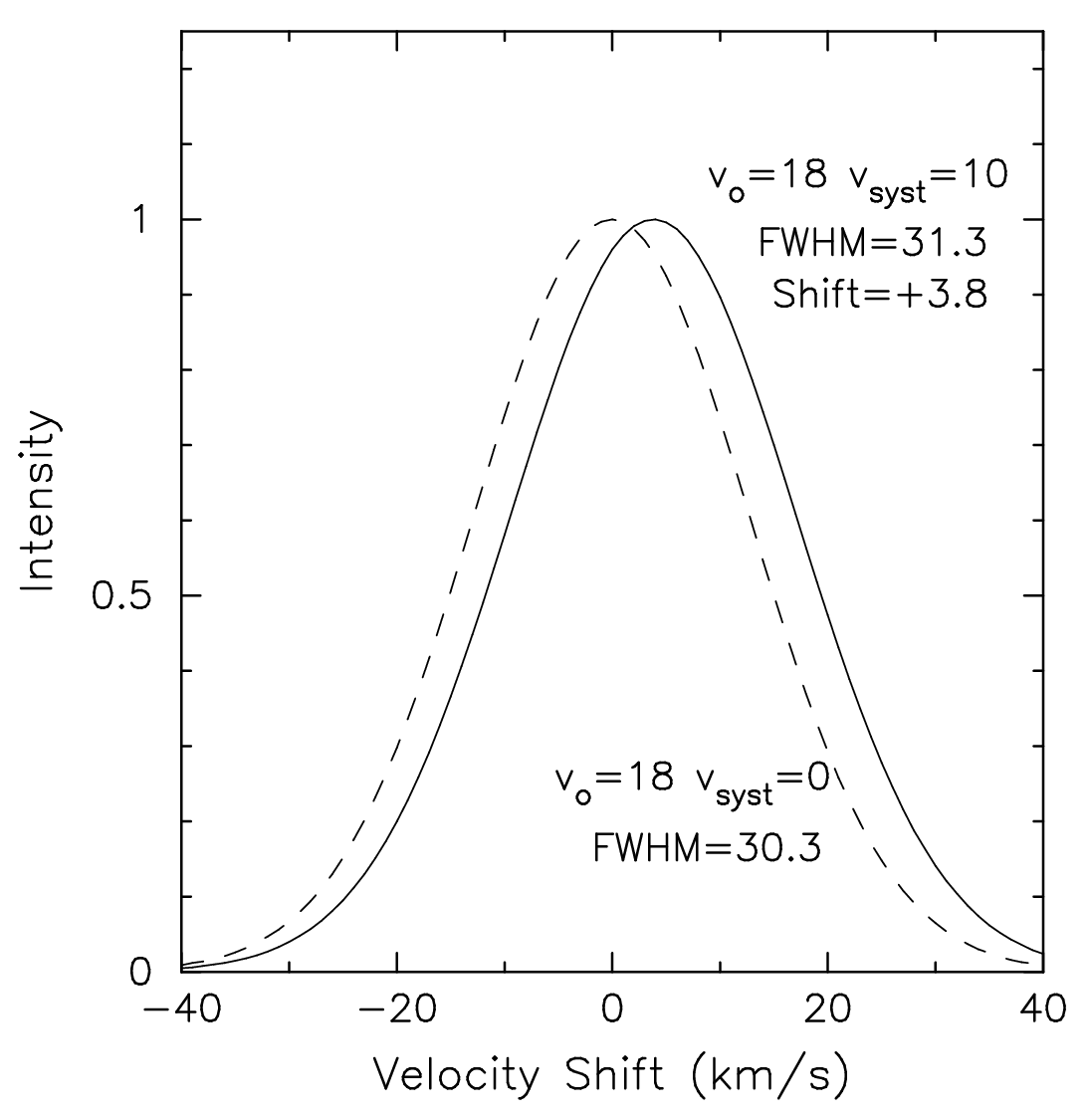}
\end{center}
\caption{Line profiles for the spherical model for two typical ions.  The dashed curve 
shows the calculated profile for a line like \Al2] excited throughout the chromosphere with 
an isotropic turbulence of 18 \kmps\ superimposed on the expansion of the chromosphere/wind.
This profile is fit to within the resolution of the plot by a Gaussian with FWHM = 30.3 \kmps.
The solid curve shows the effect of superimposing a global 10~\kmps\ downdraft on this profile.
It is again fit with a Gaussian (no \C2]-like wings) but shifted 3.8 \kmps\ to the red and 
broadened slightly to FWHM = 31.3 \kmps.  
\label{fig7}}
\end{figure}

%%Figure 8--Effect of turbulence on profiles
%%Panels a--e come from winds/analysis/anisotropic; panel f from winds/analysis/c2

\begin{figure}
\begin{center}
\epsscale{0.80}
\plottwo{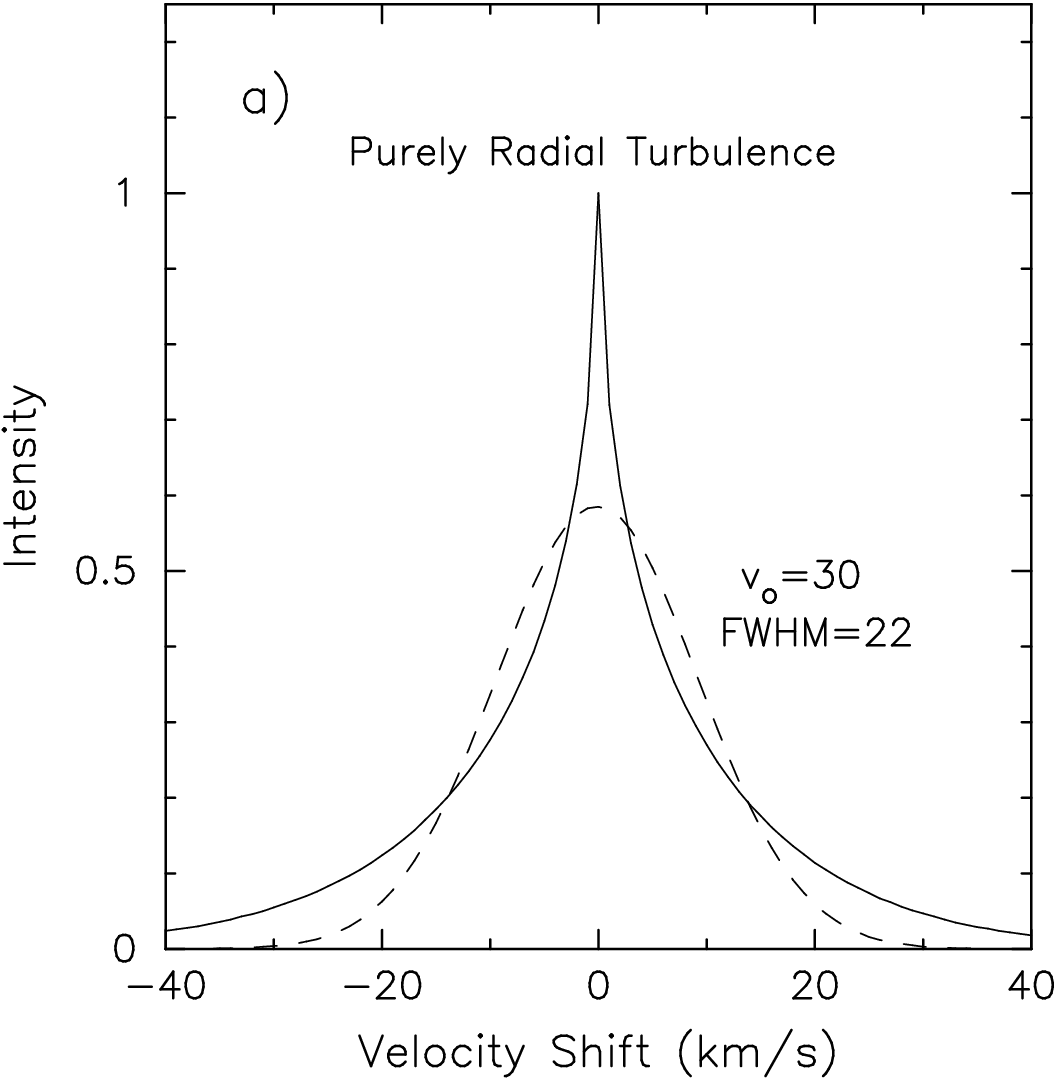}{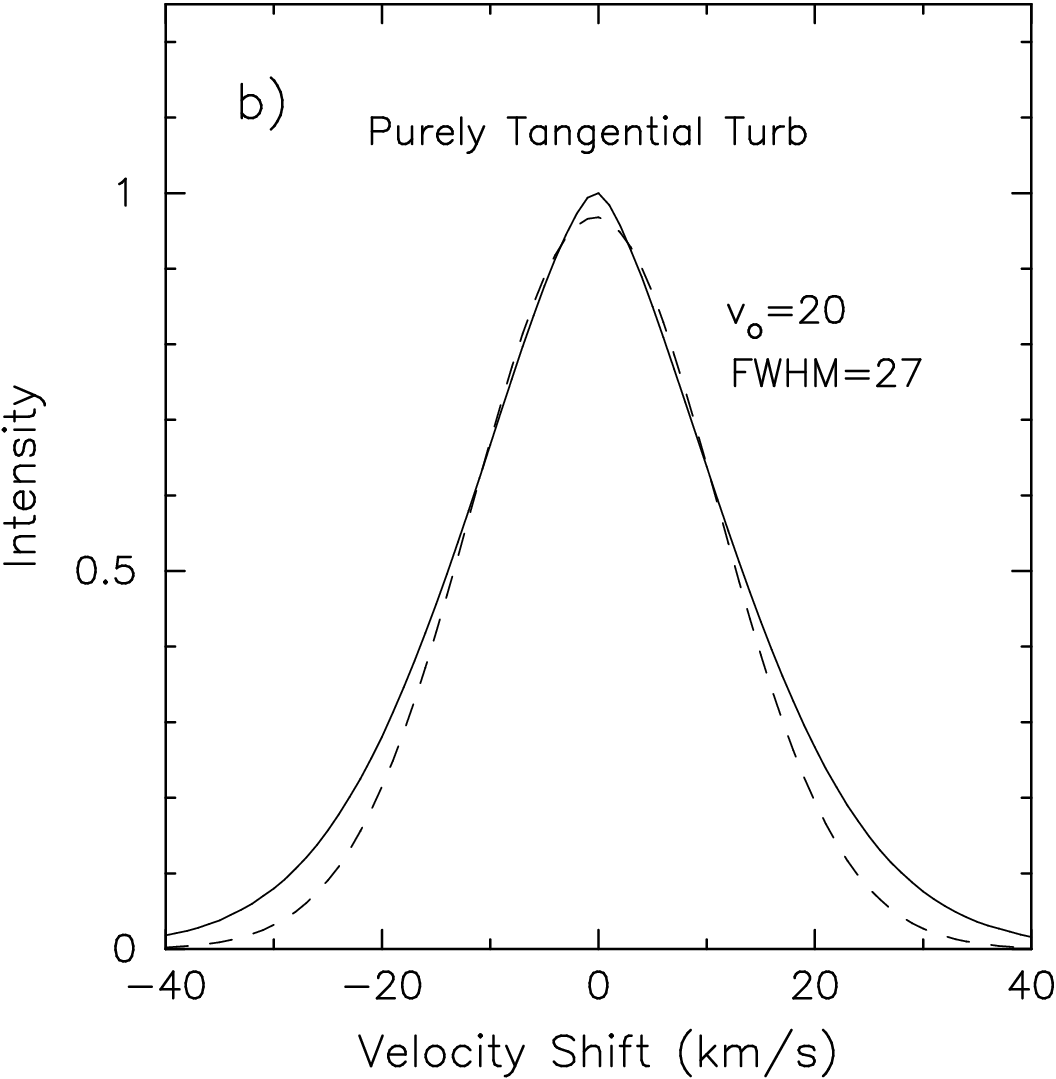}
\plottwo{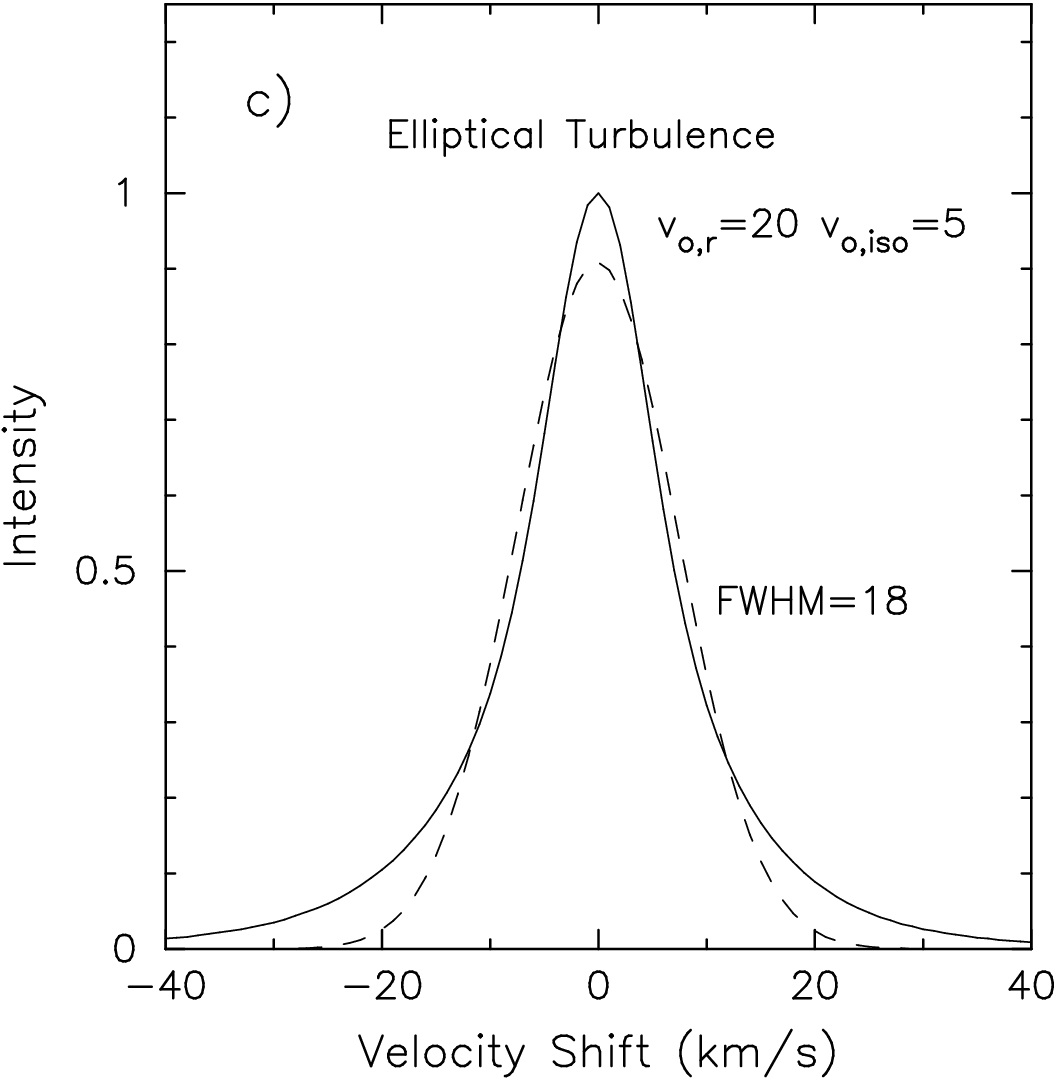}{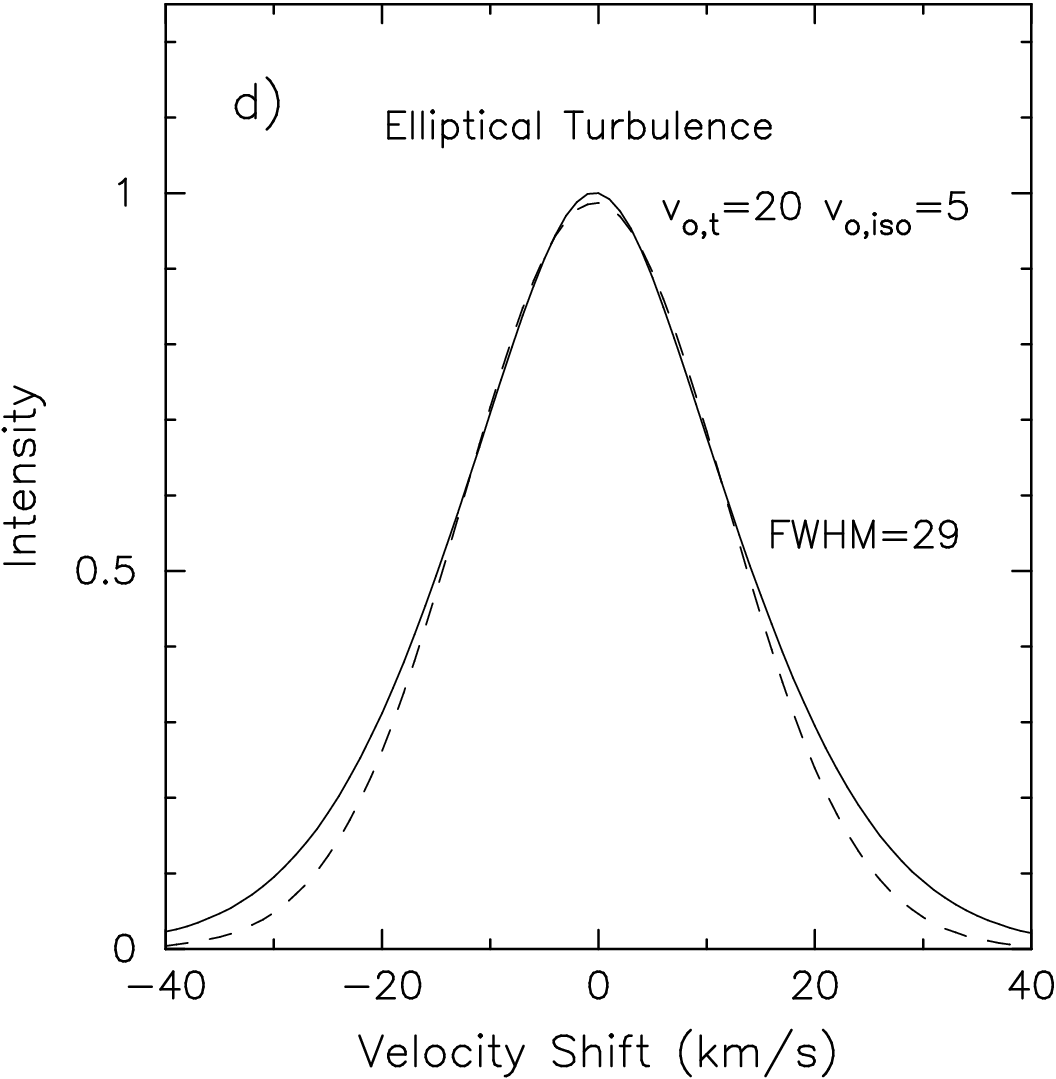}
\plottwo{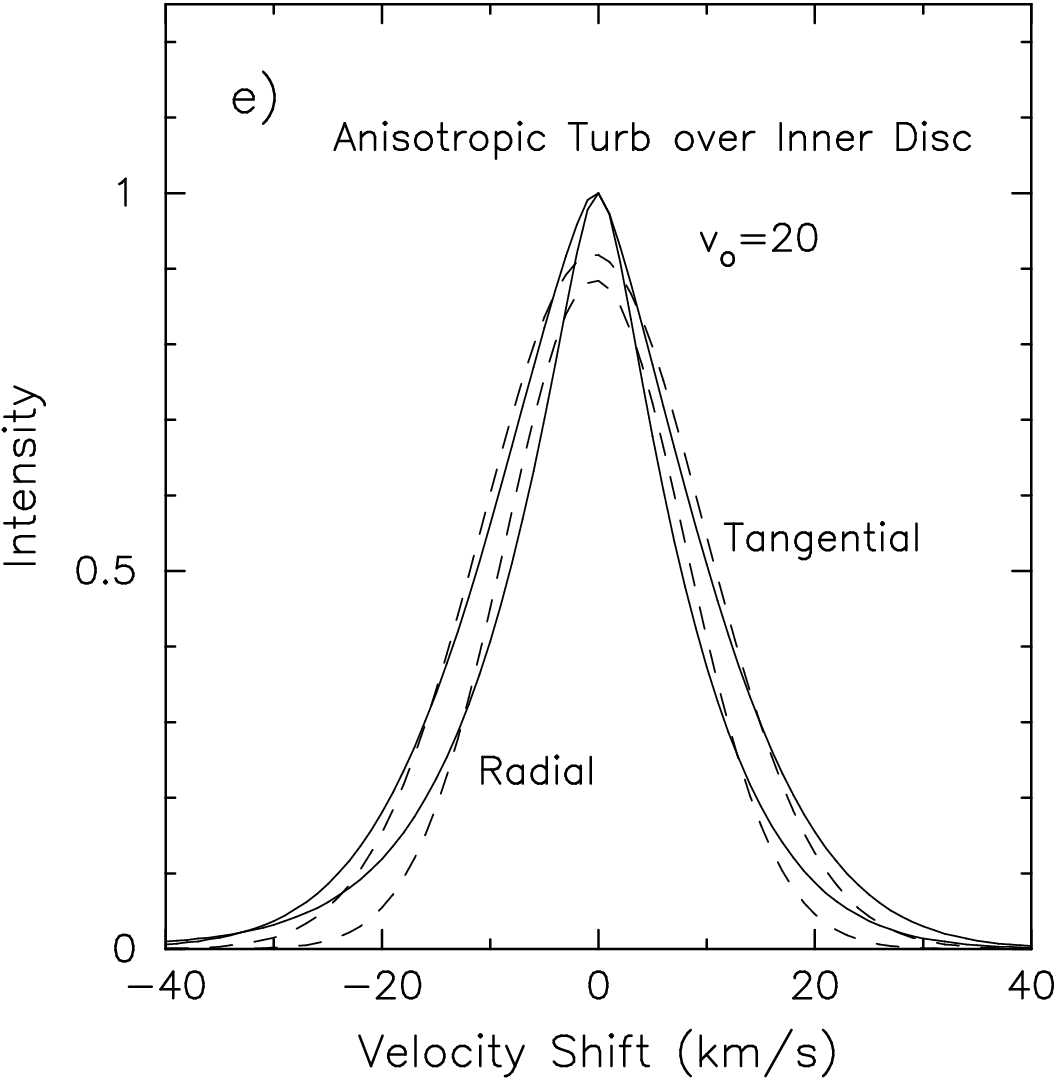}{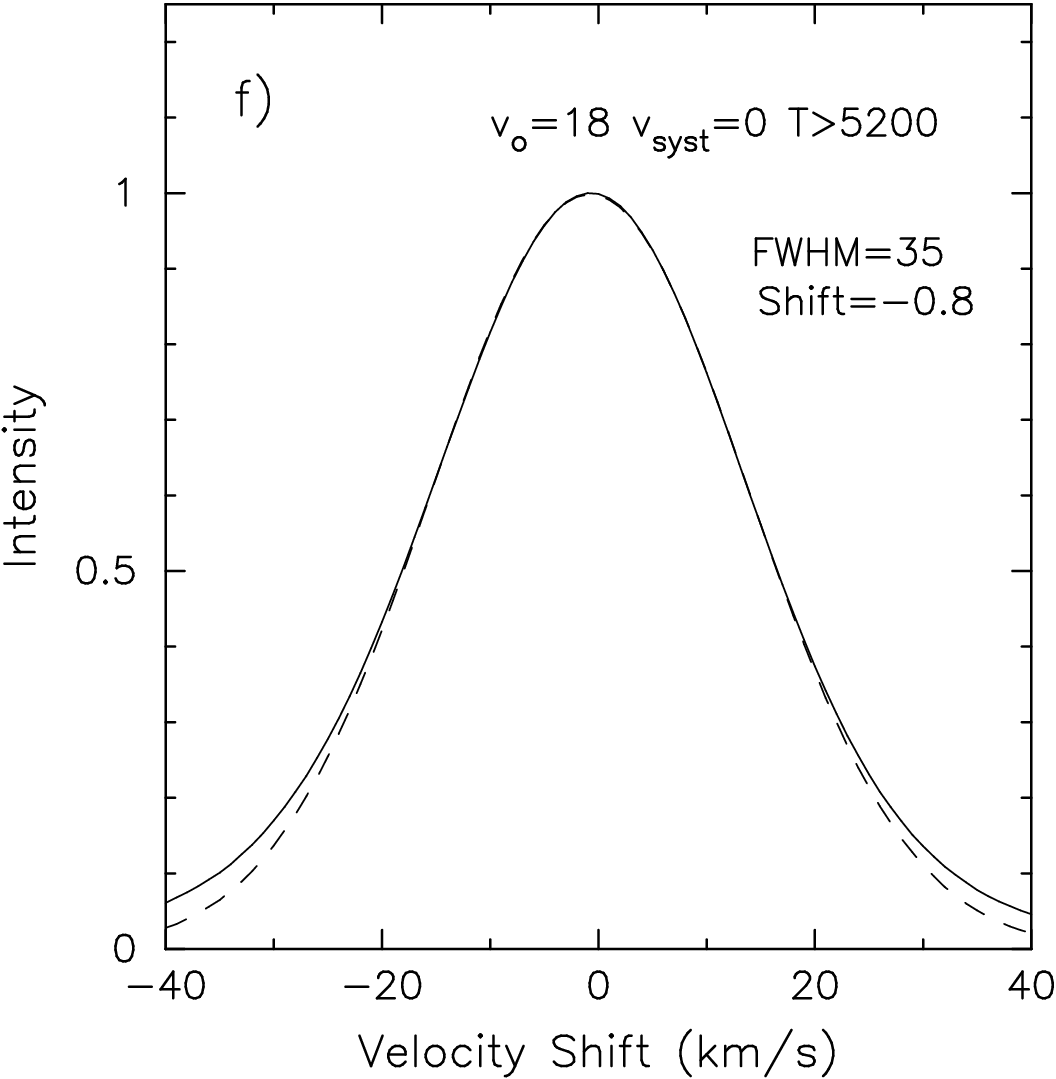}
\end{center}
\caption{\scriptsize The effects of various types of anisotropic turbulence on calculated line 
profiles.  In these figures, the solid curve is the calculated profile, and the dashed curve is 
a Gaussian fit.  For the more extreme profiles to the left, this Gaussian is fit to the whole 
profile; for the subtler profiles to the right, the Gaussian is fit to inner part 
(Intensity$\geq$0.5).  Panel a) shows the really awful effect of purely {\it radial} turbulence, 
while Panel b) shows the more subtle effect of purely {\it tangential} turbulence.  Panels c) and 
d) show, respectively, the effect of 20 \kmps\ of radial or tangential turbulence combined with 
5 \kmps\ of isotropic turbulence.  Panel e) is an experiment with confining radial and tangential 
turbulence to only the disc.  The radial distribution is noticably different than Panel a) because 
most of the emission in Panel a) comes from {\it beyond} the edge of the disc.  Panel f) shows the 
effect of isotropic turbulence with only the gas hotter than 5200~K emitting. 
\label{fig8}}
\end{figure}

%%Figure 9--Profiles pf C II] 
%%The two figures here come from winds/analysis/composite

\begin{figure}
\begin{center}
\epsscale{1.00}
\plottwo{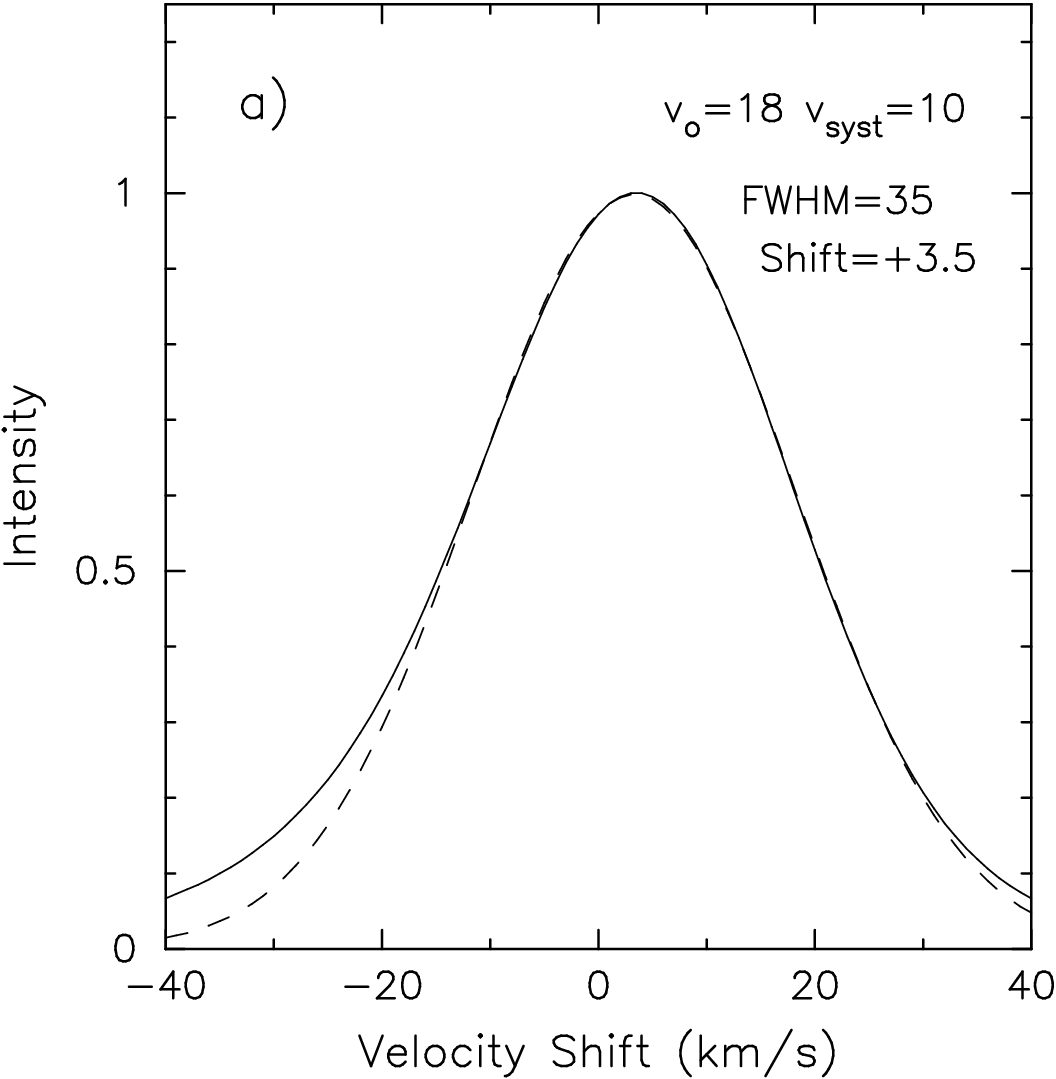}{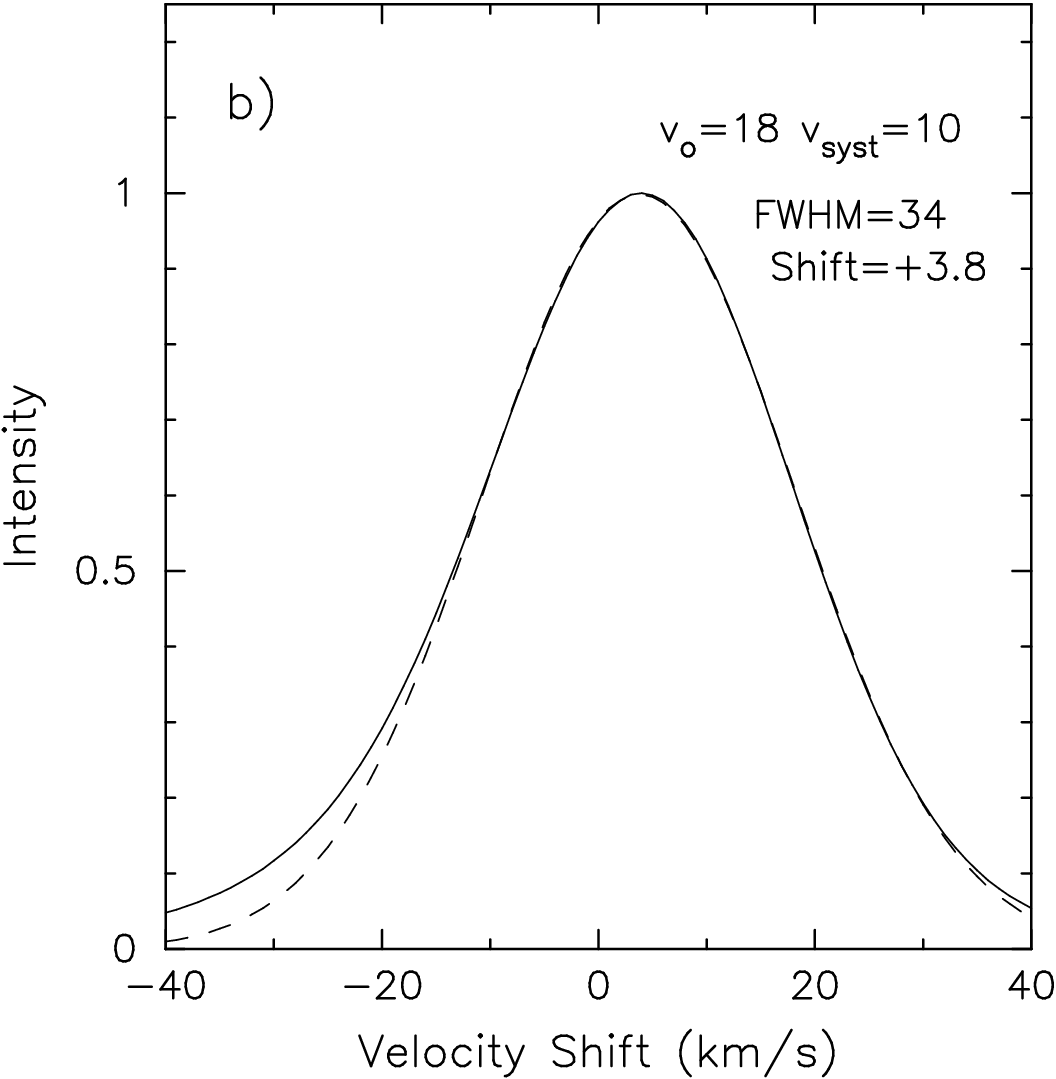}
\end{center}
\caption{An attempt to fit the peculiar shapes and shifts of \C2] emission in windy giants.
Calculated profiles are the solid curves; Gaussians fit to the inner parts of these profiles 
(I$>$0.5) are dashed curves.  In both cases gas in the deeper parts of the chromosphere 
(cooler than some threshold value) was given a 10~\kmps\ downward velocity.
Panel a) at left shows the effect of suppressing all the emission at temperatures between 
6000~K and 8500~K while letting 30\% of the gas below 6000~K emit.  Panel b) at right shows 
the effect suppressing all emission from gas cooler than 8000~K except for gas with electron 
densities above log(\nelect) $\geq$\ 8.6.  In both cases extra broadening in the wings 
comes from emission from gas with significant expansion velocity projected into the line 
of sight.  
\label{fig9}}
\end{figure}

%%Figure 10--Shell velocities in lam Vel
%%This figure comes from winds/analysis/fe2

\begin{figure}
\epsscale{0.75}
\plotone{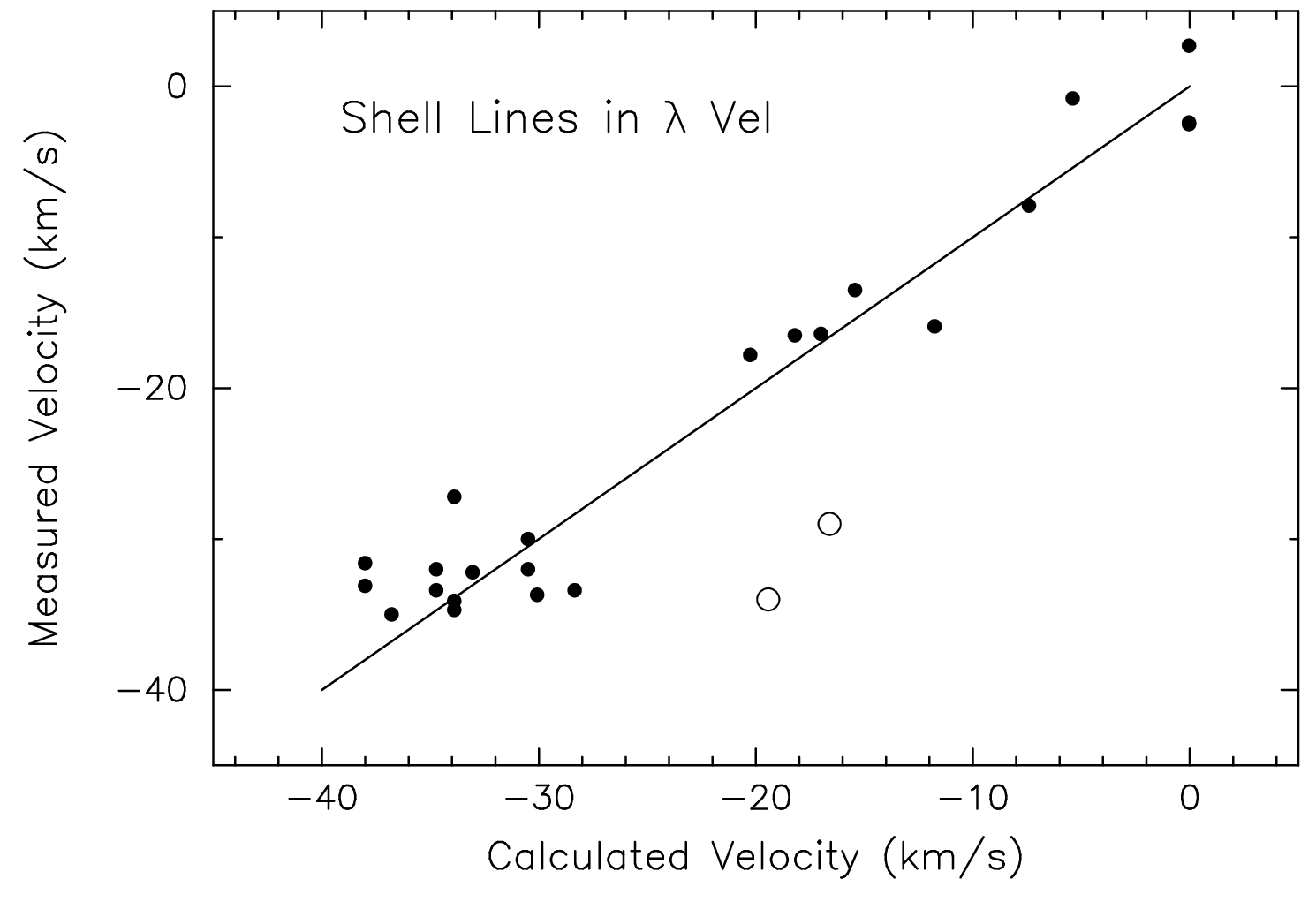}
\caption{A comparison of shell velocities for $\lambda$~Vel with calculations for the 
31~Cyg model under assumptions about mass loss and ionization.  Measured values come 
from Carpenter et al.\ (1999, Table 3).  The circles are two lines for which the measured 
values seemed discrepant in Carpenter's paper, one even falling beyond the limits of 
his Fig.\ 9.  Calculated velocities assume \Mdot = 9$\times$10$^{-9}$ \MLrate and that 
Fe is doubly ionized at \nelect\ $<$ 2$\times$10$^6$.  
\label{fig10}}
\end{figure}

\end{document}